\documentclass[aps,pra,groupedaddress,twocolumn,nofootinbib,superscriptaddress,longbibliography]{revtex4-1}
\usepackage{times}
\usepackage{amsmath}
\usepackage{amssymb,bm}
\usepackage{amsthm,color,dsfont}
\usepackage[table]{xcolor}
\usepackage{tabularx}
\usepackage{braket}
\usepackage{amssymb}
\usepackage{subcaption}
\usepackage[colorlinks=true, linkcolor=blue, urlcolor=blue, citecolor=blue, hyperindex, breaklinks]{hyperref}
\usepackage{graphicx}
\usepackage{hyperref}
\hypersetup{colorlinks=true, citecolor=blue, urlcolor=blue, linkcolor=blue}

\usepackage{ulem}
\usepackage{tikz}
\usetikzlibrary{lindenmayersystems}
\usetikzlibrary{positioning, shapes}
\newtheorem*{definition*}{Definition}

\newcommand\add[1]{{#1}}

\begin{document}

\title{Quantum circuit optimizations for NISQ architectures}

\author{Beatrice Nash}
\email[Electronic address: ]{beatricenash@fas.harvard.edu}
\affiliation{Department of Physics, Massachusetts Institute of Technology, 
Cambridge, MA, USA}
\affiliation{Department of Computer Science, Harvard University, 
Cambridge, MA, USA}
\affiliation{Institute for Quantum Computing, University of Waterloo, 
Waterloo, ON, N2L 3G1, Canada}

\author{Vlad Gheorghiu}
\email[Electronic address: ]{vlad.gheorghiu@uwaterloo.ca}
\affiliation{Institute for Quantum Computing, University of Waterloo, 
Waterloo, ON, N2L 3G1, Canada}
\affiliation{Department of Combinatorics \& Optimization, University of Waterloo, Waterloo, ON, N2L 3G1, Canada}
\affiliation{softwareQ Inc., Kitchener, ON, Canada}

\author{Michele Mosca}
\affiliation{Institute for Quantum Computing, University of Waterloo,
Waterloo, ON, N2L 3G1, Canada}
\affiliation{Department of Combinatorics \& Optimization, University of Waterloo, Waterloo, ON, N2L 3G1, Canada}
\affiliation{Perimeter Institute for Theoretical Physics, Waterloo, ON, N2L 6B9, Canada}
\affiliation{Canadian Institute for Advanced Research, Toronto, ON,  M5G 1Z8, Canada}

\date{Version of \today}

\begin{abstract}
Currently available quantum computing hardware platforms have limited 2-qubit connectivity among their addressable qubits. In order to run a generic quantum algorithm on such a platform, one has to transform the initial logical quantum circuit describing the algorithm into an equivalent one that obeys the connectivity restrictions.

In this work we construct a circuit synthesis scheme that takes as input the qubit connectivity graph and a quantum circuit over the gate set generated by $\{\text{CNOT},R_{Z}\}$ and outputs a circuit that respects the connectivity of the device. As a concrete application, we apply our techniques to Google's Bristlecone 72-qubit quantum chip connectivity, IBM's Tokyo 20-qubit quantum chip connectivity, and Rigetti's Acorn 19-qubit quantum chip connectivity. In addition, we also compare the performance of our scheme as a function of sparseness of randomly generated quantum circuits, \add{and discuss how to apply our techniques as a subroutine for the more general mapping problem over universal set of gates (Clifford + T).}
\end{abstract}

\maketitle

\section{Introduction\label{sct:intro}} 
Near-term quantum devices, such as Noisy Intermediate Scale Quantum Computers (NISQ)~\cite{Preskill2018quantumcomputingin}, are limited by sparse qubit connectivity, and many current compilers require that the input circuit takes into account the allowed connectivity of the hardware.  In this paper we introduce a circuit synthesis method that takes as input the connectivity of the hardware and the desired transformation that can be produced using gates from the set $\{\text{CNOT},R_{Z}\}$, where $R_Z$ denotes an arbitrary rotation about the $Z$ axis of the Block sphere and outputs a circuit that respects the connectivity of the device.  Our results show a significant decrease in CNOT count compared to circuit synthesis methods currently in use and allows for efficient circuit synthesis given any arbitrary connectivity. The method we present is a heuristic: finding the exact optimal solution appears intractable \cite{herr}.

The concept behind our approach is to take circuit synthesis methods that optimize the CNOT count of the output circuit that perform well under the assumption of full connectivity (\cite{patel} for CNOT circuits and \cite{6899791} for CNOT+$R_{Z}$ circuits) and modify them to take into account connectivity constraints.  We compare the results to first synthesizing the circuit using the original methods and then accounting for the constraints and found that our approach of considering connectivity constraints and synthesizing the circuit simultaneously produced sizable reductions.  These reductions depend on the sparseness of the connectivity and the complexity of the input transformation, as shown in our Results section.

Our method is effective and simple to implement, taking as input the desired transformation (the exact form of which is described in detail in the Section~\ref{sct:methods} - Methods) and the graph representing the connectivity of the device and outputting a circuit that respects the allowed connectivity.

The reminder of this paper is organized as follows. In Sec.~\ref{sct:methods} we describe our methodology, followed by our results in Sec.~\ref{sct:results}. \add{In Sec.~\ref{sct:ugs} we show how to apply our techniques to arbitrary circuits composed of gates from a universal set, discuss the limitations and also compare with state-of-the art methods of general compilers such as the one implemented in Qiskit~\cite{Qiskit}}. In Sec.~\ref{sct:conclusions} we conclude our manuscript and raise a series of open questions. A fully worked out example that illustrates our methods is depicted in Appendix~\ref{apdx:example}.

Note: Recently, the authors of~\cite{1904.00633} independently presented a similar optimization scheme.
Our work is independent of~\cite{1904.00633}, being a longer version of the seminar presented by Beatrice Nash at the Dagstuhl Seminar 18381: Quantum Programming Languages, pg. 120, September 2018, Dagstuhl, Germany~\cite{mosca_et_al:DR:2019:10329}, slide deck available online at \url{https://materials.dagstuhl.de/files/18/18381/18381.BeatriceNash.Slides.pdf}. 

\section{Methods\label{sct:methods}}

\begin{figure}[h]
\begin{subfigure}[a]{0.5\textwidth}
	\centering
	\begin{tikzpicture}
	
	\matrix[nodes={draw},
	        row sep=1cm,column sep=1cm]{
	        \node[circle,inner sep = 2pt,minimum size=1pt] (1) {$q_{1}$}; &
	        \node[circle,inner sep = 2pt,minimum size=1pt] (2) {$q_{2}$}; &
	        \node[circle,inner sep = 2pt,minimum size=1pt] (3) {$q_{3}$}; &
	        \node[circle,inner sep = 2pt,minimum size=1pt] (4) {$q_{4}$}; \\
	        };
	        \path[-] (1) edge (2);
	        \path[-] (2) edge (3);
	         \path[-] (3) edge (4);

	\end{tikzpicture}
	\caption{}
\end{subfigure}
\begin{subfigure}[b]{0.5\textwidth}
	\centering
	\includegraphics{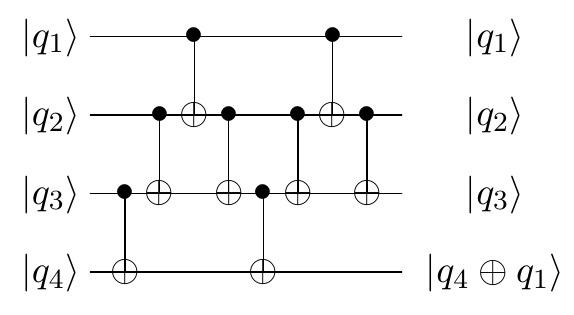}
	\caption{}
\end{subfigure}

\begin{subfigure}[c]{0.5\textwidth}
	\centering
	\includegraphics{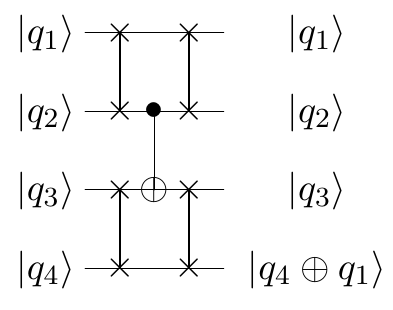}
	\caption{}
\end{subfigure}
\caption{a) 4-qubit linear nearest neighbor physical qubit connectivity. b), c) Circuits for performing a CNOT operation with $q_{1}$ as the control and $q_{4}$ as the target using the connectivity shown in a).}
\label{fgr1}
\end{figure}
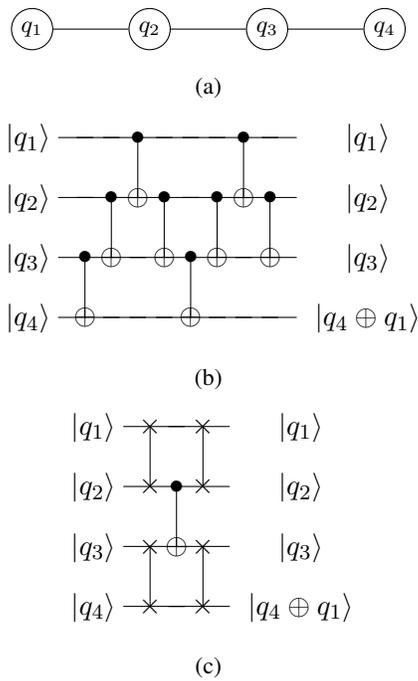

 In order to perform a CNOT operation between logical qubits mapped to non-adjacent qubits, a sequence of CNOT gates between adjacent qubits is required to indirectly perform the desired operation.  If we have linear nearest-neighbor connectivity, then to perform a CNOT between $q_{1}$ and $q_{4}$ can be achieved using the example circuits shown in Fig.~\ref{fgr1}.

These general templates can be extended to work for any physical qubit connectivity graph.  While performing a CNOT operation between any two qubits is possible, it is expensive.  The circuit depicted in Fig.~\ref{fgr1} (b) requires $4(l-1)$ CNOT gates to perform the operation, where $l$ is the distance between the two qubits. 
The na\"ive swap circuit in Fig.~\ref{fgr1} (c) requires $1+ 6(l-1)$ gates.  For circuits that dynamically reassign physical and logical qubits, the last two swap gates in Fig.~\ref{fgr1} (c) can be removed.

The best current circuit synthesis methods do not account for qubit connectivity when determining the output circuit.  CNOT operations in the optimized output circuit must be replaced by templates such as those shown previously to account for the connectivity of the physical device.  Given the sparsity of connectivity graphs of near-term quantum devices, this step increases the CNOT count drastically.

\subsection{Linear reversible circuit synthesis\label{sbsct:revs}}
The first class of circuits we look at synthesizing are linear reversible circuits consisting of only CNOT gates.  In the Clifford + T universal gate set, the only two-qubit gate needed to achieve universal quantum computation is the CNOT gate, and therefore the efficient synthesis of CNOT circuits is useful for optimizing broader classes of circuits given restricted connectivity.  Additionally, CNOT gates are far noisier than single qubit gates, and thus minimizing the number of CNOT gates in a circuit is helpful for reducing errors.

\begin{figure*}
\begin{subfigure}[c]{0.7\textwidth}
\centering
$
\setlength\arraycolsep{3pt} \begin{pmatrix}
1 & 0 & 0 & 1 \\
0 & 1 & 0 & 0  \\
0 & 0 & 1 & 0  \\
0 & 0 & 0 & 1 
\end{pmatrix}
\times
\setlength\arraycolsep{3pt} \begin{pmatrix}
1 & 0 & 0 & 0  \\
0 & 1 & 0 & 0  \\
0 & 0 & 1 & 0  \\
0 & 0 & 1 & 1 
\end{pmatrix}
\times
\setlength\arraycolsep{3pt} \begin{pmatrix}
1 & 0 & 0 & 0  \\
1 & 1 & 0 & 0  \\
0 & 0 & 1 & 0  \\
0 & 0 & 0 & 1 
\end{pmatrix}
=
\setlength\arraycolsep{3pt} \begin{pmatrix}
1 & 0 & 1 & 1  \\
1 & 1 & 0 & 0  \\
0 & 0 & 1 & 0  \\
0 & 0 & 1 & 1 
\end{pmatrix}
$
\caption{}
\end{subfigure}
\begin{subfigure}[c]{0.22\textwidth}
	\centering
	\includegraphics[scale=0.77]{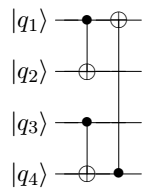}
	\caption{}
\end{subfigure}
\caption{a) Matrix operations corresponding to CNOTs in circuit in b).}
\label{fgr2}
\end{figure*}

Patel, Markov and Hayes give in \cite{patel} an asymptotically optimal algorithm for synthesizing linear reversible circuits assuming full connectivity.  The result is a $O(\frac{n^{2}}{\log n})$ size circuit in the worst case, where $n$ is the number of qubits.  In \cite{shende} it is proven that, for $n$ qubits, there exists a linear transformation for which the optimal circuit producing that transformation (again, assuming full connectivity) is size $\Omega(\frac{n^{2}}{\log_{2} n})$; hence, the method in \cite{patel} is optimal in the worst case to within a multiplicative constant.  However, each single CNOT gate in the output circuit using method \cite{patel} becomes $O(n)$ gates using the template in Fig 1. b to account for connectivity constraints.  Thus, the resulting circuit increases to size $O(\frac{n^{3}}{\log n})$ in the worst-case under connectivity constraints.  The algorithm we propose in this section improves on this performance, achieving an $O(n^{2})$ worst-case bound on the size of the resulting circuit, regardless of connectivity.

The algorithm in \cite{patel}, which is the foundation for that proposed here, takes as input an arbitrary linear transformation represented by a $n \times n$ binary matrix and outputs a circuit that produces the desired transformation.

The initial state of the $n$ qubits is represented by the $n \times n$ identity matrix and each CNOT applied produces a row operation.  Specifically, a CNOT with target $i$ and control $j$ multiplies the matrix representation of the circuit by the elementary matrix $A_{i,j}$, which is the matrix with all elements on the diagonal and $(i,j)$ equal to $1$ and all others equal to $0$.  This results in the bitwise addition of row $i$ to row $j$.  Hence, each row corresponds to the parity of the associated qubit.  An example is given in Fig.~\ref{fgr2}.

The idea behind the process is to reverse engineer a circuit from the matrix representation of the transformation.  The algorithm from \cite{patel} is an optimized version of the Gaussian elimination approach to synthesizing the circuit.  The steps are as follows:
\begin{enumerate}
\item Reduce the matrix to upper-triangular form.  Each row operation corresponds to a CNOT in the output circuit.
\item Transpose the resulting matrix and repeat, resulting in the identity matrix.
\item Construct the output circuit from the operations performed in the following order: first, the operations done in 2) with their control/targets flipped and in the same order in which they were performed, and second, the operations done 1), with their control/targets preserved but the order in which they were performed flipped.
\end{enumerate}

The $O(\frac{n^{2}}{\log n})$ upper bound on the number of operations required for this process, as opposed to the $O(n^{2})$ operations required for row reduction via Gaussian elimination, is achieved by partitioning.  The matrix is divided into $\frac{n}{\log_{2} n}$ sections of size $\log_{2} n \times n$.  Starting with the first section--after placing a $1$ on the diagonal, if necessary--eliminate duplicate sub-rows within that section before performing row reduction normally.  Then move on to the next section and do the same for the rows below the first $\log_{2} n$.  Continue in this way until the matrix is in upper triangular form, then transpose and repeat.

\begin{figure}
\begin{subfigure}[a]{0.5\textwidth}
\begin{center}
\small$\begin{pmatrix}
1 & 1 & 0 & 1 & 1 & 0 \\
0 & 0 & 1 & 1 & 0 & 1 \\
1 & 0 & 1 & 0 & 1 & 0 \\
1 & 1 & 0 & 1 & 0 & 0 \\
1 & 1 & 1 & 1 & 0 & 0 \\
0 & 1 & 0 & 1 & 0 & 1 
\end{pmatrix}$
\caption{}
\end{center}
\end{subfigure}

\begin{subfigure}[b]{0.5\textwidth}
	\centering
	\begin{tikzpicture}
	
	\matrix[nodes={draw},
	        row sep=1cm,column sep=1cm]{
	        \node[circle,inner sep = 2pt,minimum size=1pt] (1) {$q_{1}$}; &
	        \node[circle,inner sep = 2pt,minimum size=1pt] (2) {$q_{2}$}; &
	        \node[circle,inner sep = 2pt,minimum size=1pt] (3) {$q_{3}$}; \\
	        \node[circle,inner sep = 2pt,minimum size=1pt] (6) {$q_{6}$}; &
	        \node[circle,inner sep = 2pt,minimum size=1pt] (5) {$q_{5}$}; &
	        \node[circle,inner sep = 2pt,minimum size=1pt] (4) {$q_{4}$}; \\
	        };
	        \path[-] (1) edge (2);
	        \path[-] (2) edge (3);
	         \path[-] (3) edge (4);
	           \path[-] (4) edge (5);
	        \path[-] (2) edge (5);
	        \path[-] (1) edge (6);
	        \path[-] (5) edge (6);

	\end{tikzpicture}
	\caption{}
\end{subfigure}

\caption{a) Matrix representation of linear transformation; b) Connectivity of physical qubits.}
\label{fgr3}
\end{figure}
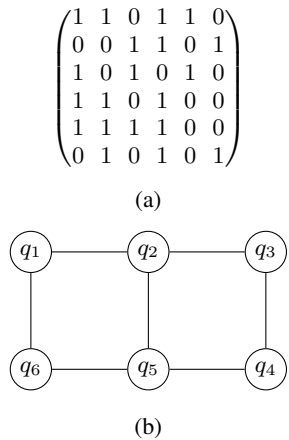

The algorithm we propose does not make use of the partitioning (although it can be easily altered to do so), because it takes advantage of grouping multiple row operations to perform together.  We found that while partitioning works well when physical qubits are fully or very nearly fully connected ($\gtrapprox$ $85$\%), for connectivity as sparse as that of near-term devices our algorithm works best without the use of partitioning.  Therefore, we do not go into more details of the process; for more information, see \cite{patel}.

Now, we give the intuition for the motivation for our method, using an example of a linear transformation on $6$ qubits and their connectivity given in Fig.~\ref{fgr3} (see Appendix~\ref{apdx:example} for the fully worked-out example). To eliminate the ones in the first column below the first row, the method in \cite{patel} uses row $1$ as the control for each of the operations.  Thus, using this example, the sequence of operations for the first column--represented as $(\text{control},\text{target})$--is $(1,3),(1,4),(1,5)$. Under the assumption of full connectivity, this makes no difference.  However, with restricted connectivity, each operation requires $4(l-1)$ CNOT gates when using the template in Fig 1.b, where $l > 1$ is the length of the shortest path between the two physical qubits.  Thus, this sequence of operations requires $4 + 8 + 4 = 16$ CNOT gates.  However, if we are to instead perform the row operations $(4,5),(3,4),(1,3)$, the ones below entry $(1,1)$ are still eliminated, and instead of $16$ CNOT gates, only $1 + 1 + 4 = 6$ are required. 

Give a connectivity graph and a set of rows, our goal is to find the shortest set of paths through the graph that effectively hits each of the nodes associated with those rows.  Then, we want to convert that path into a sequence of operations that effectively eliminates each of the rows, while leaving the rest unchanged.

\subsection{Steiner tree problem reduction\label{sbsct:steiner}}

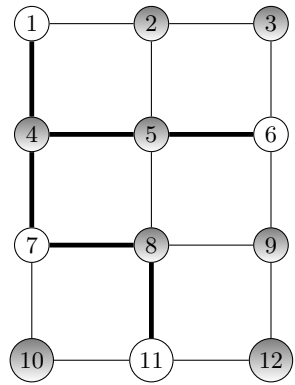
\begin{figure}
	\centering
	\begin{tikzpicture}
	
	\matrix[nodes={draw},
	        row sep=1cm,column sep=1cm]{
	        \node[circle,inner sep = 2pt,minimum size=1pt] (1) {$1$}; &
	        \node[circle,inner sep = 2pt,minimum size=1pt,shade] (2) {$2$}; &
	        \node[circle,inner sep = 2pt,minimum size=1pt,shade] (3) {$3$}; \\
	        \node[circle,inner sep = 2pt,minimum size=1pt,shade] (6) {$4$}; &
	        \node[circle,inner sep = 2pt,minimum size=1pt,shade] (5) {$5$}; &
	        \node[circle,inner sep = 2pt,minimum size=1pt] (4) {$6$}; \\
	        \node[circle,inner sep = 2pt,minimum size=1pt] (7) {$7$}; &
	        \node[circle,inner sep = 2pt,minimum size=1pt,shade] (8) {$8$}; &
	        \node[circle,inner sep = 2pt,minimum size=1pt,shade] (9) {$9$}; \\
	        \node[circle,inner sep = 2pt,minimum size=1pt,shade] (10) {$10$}; &
	        \node[circle,inner sep = 2pt,minimum size=1pt] (11) {$11$}; &
	        \node[circle,inner sep = 2pt,minimum size=1pt,shade] (12) {$12$}; \\
	        };
	        \path[-] (1) edge (2);
	        \path[-] (2) edge (3);
	         \path[-] (3) edge (4);
	           \path[-,line width = 1.8pt] (4) edge (5);
	        \path[-] (2) edge (5);
	        \path[-,line width = 1.8pt] (1) edge (6);
	        \path[-,line width = 1.8pt ] (5) edge (6);
	        \path[-,line width = 1.8pt] (6) edge (7);
	        \path[-] (5) edge (8);
	        \path[-] (4) edge (9);
	        \path[-,line width=1.8pt] (7) edge (8);
	        \path[-] (8) edge (9);
	        \path[-] (7) edge (10);
	        \path[-,line width=1.8pt] (8) edge (11);
	        \path[-] (9) edge (12);
	        \path[-] (10) edge (11);
	        \path[-] (11) edge (12);
	         
	\end{tikzpicture}
	\caption{A solution to the Steiner tree problem on this graph, with each edge having weight $1$.  The non-shaded nodes are the terminals.  The bold edges are those included in the solution.  Nodes $4$, $5$, and $8$ are Steiner nodes.}
	\label{fgr4}
\end{figure}

This problem reduces to the \textit{Steiner tree} problem on the connectivity graph $G$ with edge weights of $1$ and the set $S$ equal to the nodes associated with the control and the set of rows to be eliminated.  The Steiner tree problem is that of finding the minimum weight tree, $T$, that is a subgraph of $G$ and includes, but is not limited to including, all nodes in $S$.  Nodes in $S$ are called \textit{terminals}, and nodes in $T$ but not in $S$ are known as \textit{Steiner nodes}.  An example is shown in Fig.~\ref{fgr4}.  The reason for this reduction, perhaps not instantly clear, will become so once shown how to transform a tree into a sequence of row operations.  

We want to translate this tree, $T_{\{c,S\}}$, where $c$ is the control, into a sequence of operations that effectively eliminates each row in $S\setminus \{c\}$ using only operations between adjacent nodes.  $S \setminus \{c\}$  is the set of rows in the linear transformation matrix with ones in their $c$th entry.  We now describe in detail the process of converting a tree to a sequence of row operations.  Note that the process differs between whether the row operations are performed before or after the transpose step.

In order to convert the tree into a sequence of row operations, we first separate it into a set of edge disjoint sub-trees $\{T_{\{c_{i},S_{i}\}}\}$ with root $c_{i} \epsilon S$ and leaves $S_{i} \setminus \{c_{i}\} \subset S$.  The remaining nodes in each sub-tree are Steiner nodes.  The first sub-tree, $T_{\{c_{1},S_{1}\}}$, is rooted at $c$.  Starting from $c$, grow the sub-tree by traversing $T_{\{c,S\}}$ in breadth first search order.  When arriving at a non-leaf terminal $u$, add $u$ to $T_{\{c_{1},S_{1}\}}$ as a leaf and create a new sub-tree containing a copy of $u$ as a root.  Once the sub-tree rooted at $c$ is complete, build the sub-trees rooted at its leaves.  Continue until all the edges in $T_{\{c,S\}}$ have been added to a sub-tree.  The root of each sub-tree will be used as the control to eliminate each of its leaves.

Compute the sequence of row operations as follows.  Starting with the \textit{last} sub-tree constructed, traverse the tree in reverse depth first search order.  When traversing an edge $(u,v)$, where $u$ is closer to the root than $v$, add a row operation to the sequence with $u$ as the control and $v$ the target.  Once the top of the tree is reached, we have a sequence of operations $R$.  Let $R' = reverse(R - R[j])$, where $R[j]$ is the last operation in $R$.  Now, create $R^{*}$ by removing from $R+R'$ those operations with terminals as the targets.  $R+R'$ applies the row eliminations, while $R^{*}$ undoes those performed on the Steiner nodes, leaving them unchanged.  $R+R'+R^{*}$ gives the completed sequence of operations for that sub-tree.  Add $R+R'+R^{*}$ to the overall sequence of operations and move on to the prior sub-tree until all have been traversed.

\begin{figure}[t]
	\centering
	\includegraphics[scale=0.75]{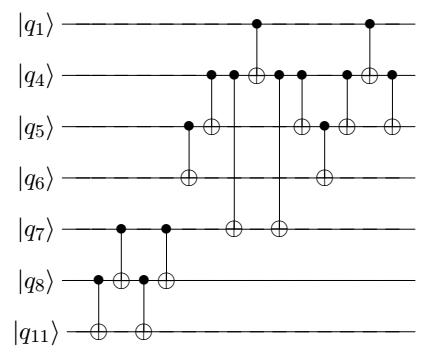}
	
	\caption{Resulting sequence of operations from the tree in Fig 4.}
	\label{fgr5}
\end{figure}

We give an example in Fig.~\ref{fgr5} for the resulting sequence of operations for the tree in Fig.~\ref{fgr4} (assuming before the transpose step), using node $1$ as the overall control.  Each CNOT in the circuit shown corresponds to a row operation with the same control and target; they are shown this way to help visualize the result.  In the resulting circuit, just as in the original algorithm, if these eliminations are performed before the matrix is transposed, their order will be flipped in the resulting circuit.  If they occur after, their control/targets will be flipped, but their order will be preserved.

Each edge will be traversed exactly once in this process.  Thus, the total number of gates required will be approximately $4*(l-d)$, where $d$ is the number of terminals (excluding the control) and $l$ is the number of edges in the tree.  It is clear that for minimum $l$, this is the optimal solution; hence the reduction to the Steiner tree problem.

The minimal Steiner tree problem, however, is NP-hard.  There are approximation algorithms that come close to optimal the best being within a factor of $1.39$, given in \cite{byrka}, which improved on the previous bound of $1.55$, given in \cite{robins}. However, their run-time is insufficient for our algorithm, which requires many iterations, two for each row.  The algorithm we use, given in \cite{wang}, is somewhat of a combination between the better-performing and more efficient approximation algorithms.

Given a set of terminals and a connectivity graph, the algorithm performs breadth-first search outwards from each of the terminals.  When the paths collide, the nodes along that path consolidate into a single node and all the edges adjacent to the consolidated nodes are placed adjacent to this new node.  The process is restarted with this node as a new terminal.  The total time will therefore be $O(d*(|V| + |E|))$, where $d$ is the number of terminals.  The resulting Steiner tree is within a factor of $2(1-\frac{1}{l})$ times the size of the optimal tree, where $l$ is the number of leaves in the optimal tree.  From many trials, it seems that this approximation is sufficient to see a large reduction in the CNOT count of the output circuit.  The choice of Steiner tree approximation algorithm for this purpose depends on the user's efficiency and performance requirements; a survey is given in \cite{hwang}.

The algorithm thus far works as follows: 
\begin{enumerate}

\item Start with column $i=1$.
\item If entry $(i,i) = 0$, find all rows $j$ such that $(j,i) = 1$ and $j > i$.  Choose $j$ with the shortest path in the connectivity graph to $i$, and use this path to perform a series of allowed row operations that adds row $j$ to row $i$.
\item Find rows below row $i$ with entry in column $i$ equal to $1$.  Create set of terminals $S$ from nodes associated with those rows in addition to node ${i}$.
\item Find Steiner graph approximation with connectivity graph $G$ and terminals $S$.
\item Compute row operations using this graph that eliminate those rows with constraints on allowed row operations.  Perform those operations and compute resulting matrix.
\item Repeat steps 2-5 on the next columns, until the matrix is in upper-triangular form.
\item Transpose the matrix and repeat.

\end{enumerate}

After the matrix is transposed, we must alter step $5$ in the process slightly.  Now, say that a row operation is effectively performed between $(i,j)$, with $i$ as the control, and $j$ as the target.  If $i > j$, then the lower-triangular form of the matrix is ruined.  Thus, we have to alter the algorithm so as to only allow effective row operations with a lower index row as the control and a higher index as the target.

\begin{figure}[b]
\centering
\includegraphics[scale=0.75]{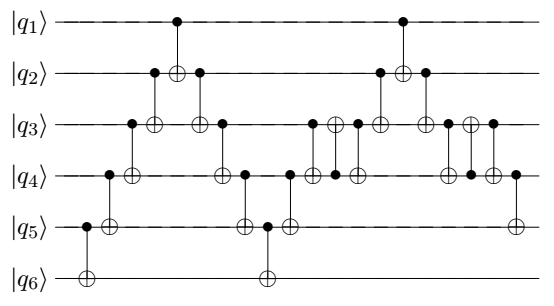}
\caption{Circuit that produces CNOT operations between both $q_{1},q_{3}$ and $q_{1},q_{6}$}
\label{fgr6}
\end{figure}

To do this, perform the Steiner tree algorithm as before, except use the smallest node as the control for all of the operations.  Fig.~\ref{fgr6} shows how to eliminate two rows on the same path, which is never necessary before the transpose step.  From Fig.~\ref{fgr6} it is clear that the maximum number of additional gates required to ensure this requirement is $4*d$.

Let the output sequence of operations before the transpose step be a list $A$ and after the transpose step be a list $B$.  When adding each operation to the output sequence $B$, flip the control and target.  The output circuit will be: $B + \textit{reverse}(A)$.

The Steiner tree approximations will always be size $O(n)$.  Since the number of operations is $O( 4*(\text{size of tree}))$, then the operations required will be $O(n)$.  Since there are $O(n)$ trees computed in total throughout the execution of the algorithm ($O(1)$ per column), then the overall number of operations is $O(n^{2})$, regardless of connectivity.

\subsection{CNOT + phase circuit synthesis\label{sbsct:CNOTphase}}

The other class of circuits we look at synthesizing given connectivity constraints are those consisting of CNOT gates and $Z$-basis rotations of arbitrary angles.  Amy, Azimzadeh, and Mosca give in \cite{6899791} a heuristic algorithm for the efficient synthesis of CNOT count optimized circuits in this class assuming full connectivity.  The results in \cite{6899791} show a $23\%$ decrease in the CNOT count on average for a suite of \textit{Clifford+T} benchmark circuits.  

The $\{CNOT,R_{z}\}$ circuit to be synthesized is described by its \textit{phase polynomial} $f$ and matrix $A$ representing the overall linear transformation on the basis states.  $(f,A)$, known as the \textit{sum-over-paths} form, fully defines the desired unitary transformation applied by the output circuit, $U_{C}$.  For a size $n$-qubit circuit:

\begin{center}
$
U_{C} = \sum_{\mathbf{x} \epsilon \mathbb{F}_{2}^{n}} e^{2\pi i f(\mathbf{x})} \ket{A \mathbf{x}}\bra{\mathbf{x}},
$
\end{center}

\noindent where $\mathbb{F}_{2}^{n}$ is the set of all length $n$ bit strings, and $f(\mathbf{x})$ is given by:

\begin{center}
$
f(\mathbf{x}) = \sum_{\mathbf{y} \epsilon \mathbb{F}_{2}^{n}} \hat{f}(\mathbf{y}) (x_{1}y_{1} \oplus x_{2}y_{2} \oplus ... \oplus x_{n}y_{n}).
$
\end{center}

\begin{figure}[t]
\centering
\includegraphics[scale=0.5]{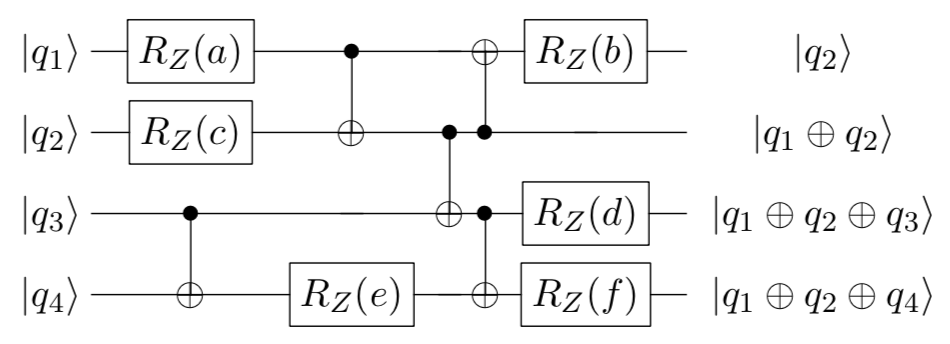}
\caption{CNOT + phase circuit example}
\label{fgr7}
\end{figure}

\noindent $f(\mathbf{x})$ is the Fourier expansion of $f$ with Fourier coefficients $\hat{f}(\mathbf{y})$.  Let the support of $\hat{f}$, $\text{supp}(\hat{f})$, be the parities with nonzero Fourier coeffiecient.   Since the application of a CNOT gate with control $x_{1}$ and target $x_{2}$ maps $\ket{x_{1}, x_{2}}$ to $\ket{x_{1}, x_{1} \oplus x_{2}}$, the state of each qubit can be mapped to a bit string representing a parity.  Each $R_{Z}$ gate is determined by the parity of the state of the qubit on which it is applied at that point in the circuit.  The coefficients for each parity $\mathbf{y}  \epsilon \mathbb{F}_{2}^{n}$, $\hat{f}(\mathbf{y})$, are given by the sum of the $Z$ rotation angles on that parity.   Hence $R_{Z}$ gates acting on the same parity can be combined.   For example, the phase polynomial $f$ and basis state transformation $A$ associated with the circuit shown in Fig.~\ref{fgr7} is:

\begin{align*}
f = &a(x_{1}) + (b+c)(x_{2}) + d(x_{1} \oplus x_{2} \oplus x_{3})
\\ &+ e(x_{3} \oplus x_{4}) + f(x_{1} \oplus x_{2} \oplus x_{4})
\end{align*}
\begin{align*}
A = \begin{pmatrix} 0 &1 &0 &0 \\
1 &1 & 0 &0 \\
1 &1 &1 &0 \\
1 &1 &0 &1
\end{pmatrix}
\end{align*}

The method in \cite{6899791} first synthesizes a circuit with the input phase polynomial $f$, then uses the method in \cite{patel} to produce the basis state transformation given by input matrix $A$.  To do the former, the goal is to compute a minimal \textit{parity network}.  A parity network is a circuit in which every parity in a set $S$ appears.  In our case, $S = \text{supp}(\hat{f})$.  By applying $R_{Z}(\theta)$, with $\theta = \hat{f}(\mathbf{y})$, to each parity $\mathbf{y} \epsilon S$, the resulting circuit will have the desired phase polynomial.  To apply the linear transformation to the basis states given by $A$, first compute the linear transformation $C$ resulting from the parity network.  Using the method described previously, compute the circuit the linear transformation $AC^{-1}$ and append it to the end of the existing circuit.

To find the exact minimal parity network, however, is NP-hard; a a heuristic for synthesizing an approximation is given in \cite{6899791}.  The algorithm works as follows:
\begin{enumerate}
\item Represent the set of parities as a matrix $P$, where each column corresponds to a parity and each row to a qubit.  For example, $P$ associated with the circuit given in Fig.~\ref{fgr7} is 
\begin{center}
$\begin{pmatrix}
1 & 0 & 1 & 1 & 0 \\
0 & 1 & 1 & 1 & 0 \\
0 & 0 & 1 & 0 & 1 \\
0 & 0 & 0 & 1 & 1
\end{pmatrix}$
\end{center}
where the first column corresponds to parity $x_{1}$, the second to parity $x_{2}$, the third to parity $x_{1} \oplus x_{2} \oplus x_{3}$, the fourth to parity $x_{1} \oplus x_{2} \oplus x_{4}$, and the last to parity $x_{3} \oplus x_{4}$.
\item Find the row $j$ (of those not yet recursed on) such that:
\begin{center}
$j = \text{arg max}_{i \epsilon \text{cols}(P)} \text{max}_{x \epsilon \{0,1\}}\{|P_{j,i} = x|\}.$
\end{center}
\item Separate $P$ into $P^{0}$, the columns $i$ with $P_{j,i} = 0$, and $P^{1}$, those with $P_{j,i} = 1$.
\item Recurse on $P^{0}$.
\item For $P^{1}$, find row $i \neq j$ such that all elements in row $i$ of $P^{1}$ equal $1$.  Add a CNOT with control $i$ and target $j$ to our parity network.  Set $P_{i} = P_{i} + P_{j}$.  Repeat until no more such rows are found.
\item Recurse on $P^{1}$.
\end{enumerate}
A detailed example of this process can be found in \cite{6899791}.  Our method performs the same series of steps, except step $5$ is modified as follows: 
\begin{enumerate}
\item[5.] Compute the approximate Steiner tree of the connectivity graph $G$ with $S$ equal to the set of rows $\{i\}$ such that each element in row $i$ of $P^{1}$ equals $1$.  Using row $j$ as the control, eliminate them all together in the same manner described in the previous section.
\end{enumerate}
In this case, whether the control bit has greater index than the target is not relevant.

Each parity in $\text{supp}(\hat{f})$ will appear at least once in this circuit.  For each $\mathbf{y } \text{ } \epsilon \text{ supp}(\hat{f})$, apply gate $R_{Z}(\hat{f}(\mathbf{y}))$ to the qubit with incoming parity $\mathbf{y}$. Lastly, append the circuit producing linear transformation $AC^{-1}$, taking into account connectivity constraints, using the algorithm described previously.

\section{Results\label{sct:results}}

In this section we compare the size of the output circuits generated using the algorithm described in the previous sections to those generated by first synthesizing the circuit using the method in~\cite{patel} (for linear reversible circuits) or~\cite{6899791} (for CNOT + T circuits) without taking into account connectivity constraints, then, once the circuit has been synthesized, using the template given in Fig.~\ref{fgr1} (b) to take into account the connectivity.  In both cases, after the circuit has been synthesized, we further optimize the size of the resulting circuit by commuting operators and canceling where possible.

\begin{figure}
\centering
\includegraphics[width=0.5\textwidth]{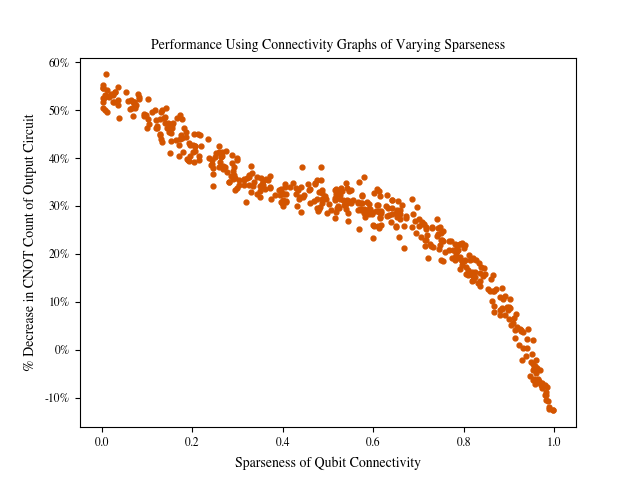}
\caption{Results for the synthesis of CNOT circuits producing a randomly generated linear transformation on 20 qubits for connectivity of varying sparseness.  We repeatedly generate random connected graphs of sparseness between 0.0 and 1.0, where sparseness is defined as the probability that an edge is placed between two qubits.  Sparseness of 1.0 is a fully connected graph.  The size of the output circuit is compared against the size of the circuit produced by first using the method in \cite{patel} (with partitioning) to synthesize a circuit assuming full connectivity and then taking into account connectivity constraints.  The reason that our method performs worse than the method in \cite{patel} for nearly complete connectivity is due to partitioning, which our method does not use, but which achieves an improvement when the qubits are nearly fully connected.  Since qubits in NISQ devices only have nearest-neighbor connectivity, their connectivity is very sparse.} 
\label{fgr8}
\end{figure}

\begin{figure}
\includegraphics[width=0.5\textwidth]{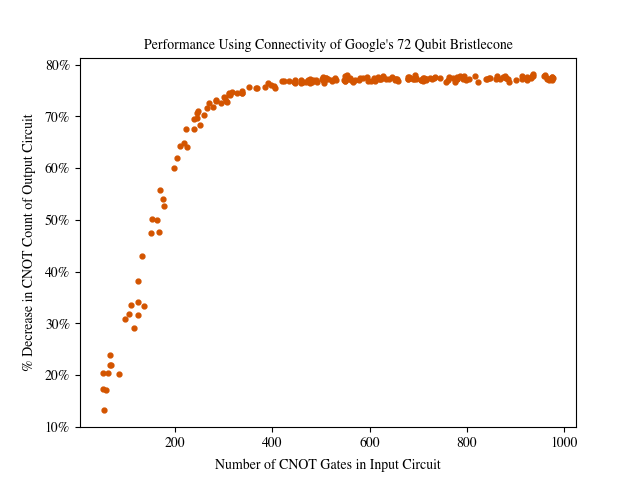}
\caption{Results for the synthesis of CNOT circuits producing randomly generated linear transformations of various size on Google's Bristlecone 72-qubit topology.  The performance is calculated as described in the caption of Fig.~\ref{fgr8}.}  
\label{fgr9}
\end{figure}

\begin{figure*}
\begin{minipage}{0.5\textwidth}
\includegraphics[width=\textwidth]{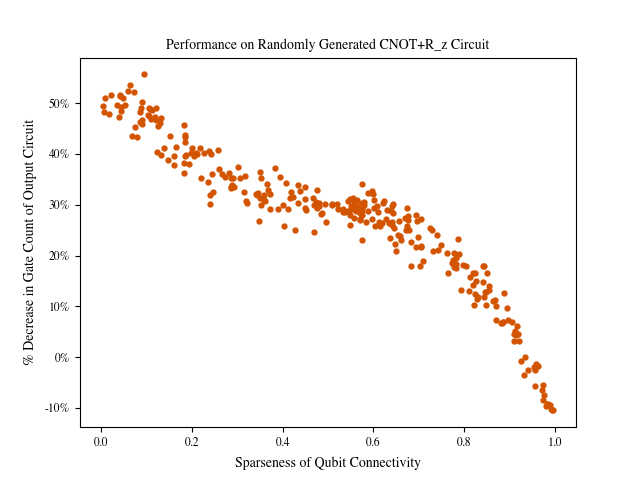}
\caption{Results for the synthesis of CNOT+$R_{Z}$ circuits producing a randomly generated linear transformation and $Z$-axis rotation on 20 qubits for connectivity of varying sparseness.  The size of the output circuit is compared against the size of the circuit produced by first using the method in \cite{6899791} to synthesize a circuit assuming full connectivity and then taking into account connectivity constraints.}
\label{fgr10}
\end{minipage}\hfill
\begin{minipage}{0.49\textwidth}
\includegraphics[width=\textwidth]{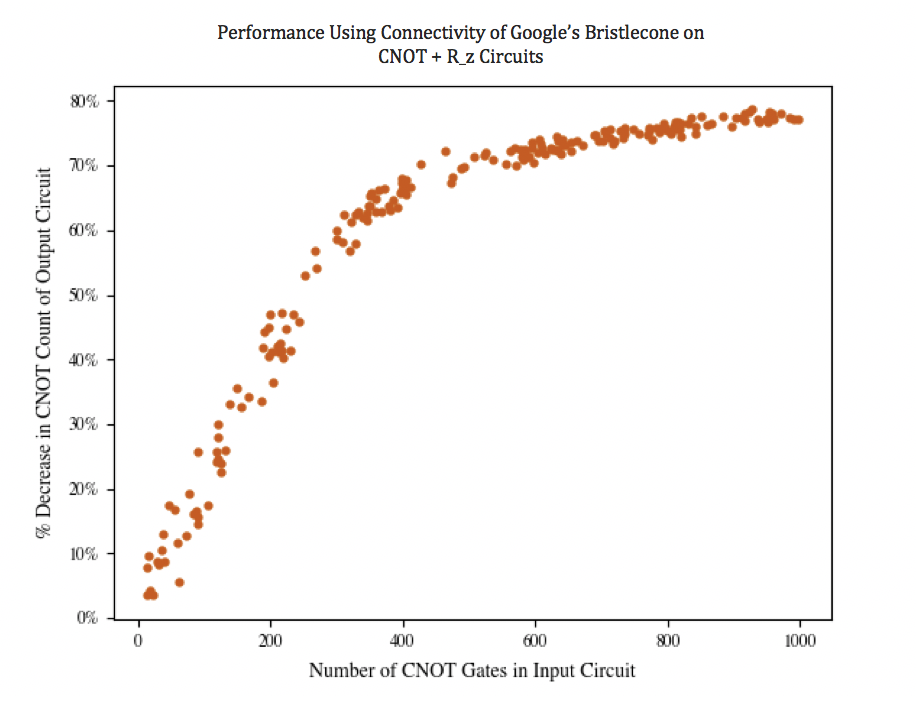}
\caption{Results for the synthesis of CNOT+$R_{Z}$ circuits producing randomly generated transformations for Google's Bristlecone 72-qubit  topology.  The size of the output circuit is compared against the size of the circuit produced by first using the method in \cite{6899791} to synthesize a circuit assuming full connectivity and then taking into account connectivity constraints.}
\label{fgr11}
\end{minipage}
\begin{minipage}{0.5\textwidth}
\includegraphics[width=\textwidth]{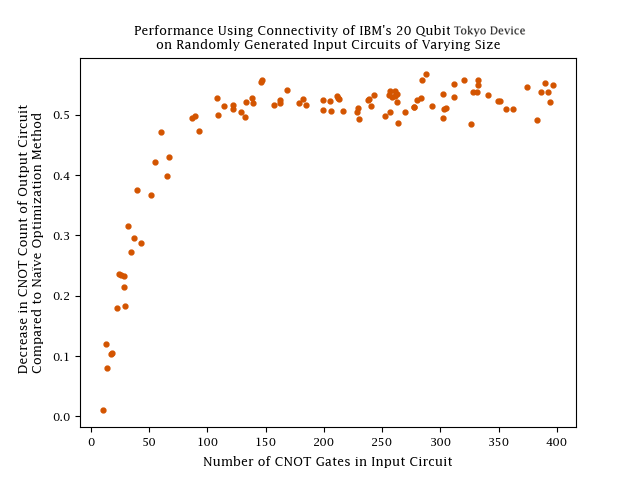}
\caption{Results for the synthesis of CNOT+$R_{Z}$ circuits producing randomly generated transformations for IBM's Tokyo 20-qubit topology.}
\label{fgr12}
\end{minipage}\hfill
\begin{minipage}{0.5\textwidth}
\includegraphics[width=\textwidth]{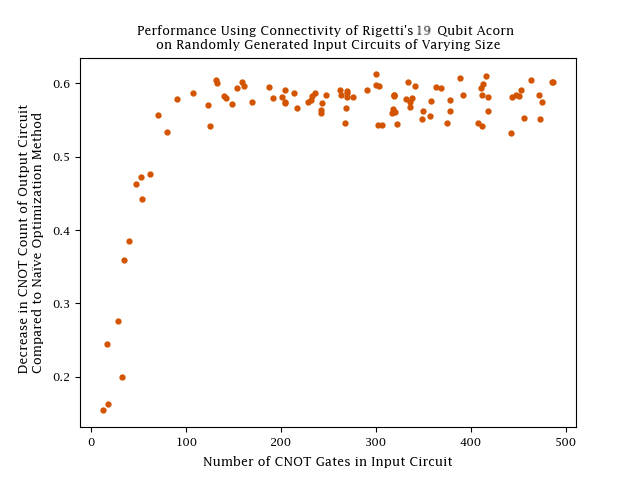}
\caption{Results for the synthesis of CNOT+$R_{Z}$ circuits producing randomly generated transformations for Rigetti's Acorn 19-qubit topology.}
\label{fgr13}
\end{minipage}
\end{figure*}

We investigate random circuits on 20 qubits with varying connectivity (results depicted in Fig.~\ref{fgr8} and Fig.~\ref{fgr10}), the Google Bristlecone 72 qubit architecture (results depicted in Fig.~\ref{fgr9} and Fig.~\ref{fgr11}), the IBM Tokyo 20 qubit architecture (results depicted in Fig.~\ref{fgr12}, and the Rigetti's Acorn 19 qubit architecture (results depicted in Fig.~\ref{fgr13}.

\add{\section{Universal Gate Sets\label{sct:ugs}}
We briefly discuss extending the applications of the method described in this paper to universal gate sets.  Circuits consisting of CNOT and phase gates alone are not sufficient for universal quantum computing. Adding the Hadamard gate, $H = \frac{1}{\sqrt{2}} \begin{pmatrix}1 & 1 \\ 1 & -1 \end{pmatrix}$, to the $\{\text{CNOT},R_{z}\}$ gate set makes it universal.  Without compromising the universality of the set, we restrict to specific $z-$rotation gates:
\begin{align*}
    S = \begin{pmatrix}1 & 0 \\ 0 & i \end{pmatrix} \;\;, T = \begin{pmatrix}1 & 0 \\ 0 & e^{\frac{i\pi}{4}} \end{pmatrix}
\end{align*}
and their conjugate transposes. Let $\mathbb{U}$ be the set of circuits composed solely of these gates.}

\add{We can apply the methods described here to this more powerful class of circuits as follows.  Given a circuit $C$ in $\mathbb{U}$, we can optimize for the number of $H$ gates in the circuit using the "merge and delete" method \cite{abdessaied14}.  Then, partition $C$ into $2*k$ segments $S_{a,b}$,  $1 \leq a \leq k$, $b \in \{\text{CNOT},H\}$, such that\begin{align} S_{k,\text{CNOT}}S_{k,H}...S_{1,\text{CNOT}} S_{1,H} = C\end{align}
and each $S_{i,H}$ contains only $H$ gates while each $S_{i,\text{CNOT}}$ contains only gates from $\{\text{CNOT},S,T,S^{\dag},T^{\dag}\}$.  Next, we can apply the methods discussed in previous sections to each of the $S_{i,\text{CNOT}}$ segments in order to account for connectivity constraints.  As discussed in the previous section, our method performs better as gate count increases, so, ideally, we want to partition the input circuit so that the $S_{i,\text{CNOT}}$ segments are large.}

\add{To do so, we use a simple heuristic technique:
first, naively partition the input circuit.  Say that the size of $C$, excluding $H$ gates, is $n$.  Then, for $i=n$ to $1$, commute gate $v_{i} \in \{\text{CNOT},S,T,S^{\dag},T^{\dag}\}$ forward through the circuit until reaching a gate with which it does not commute.  Keep track of the largest segment $S_{i_{\text{max}},\text{CNOT}}$ that $v_{i}$ can belong to.  Once reaching the end of the circuit or a non-commuting gate, insert $v_{i}$ into $S_{i_{\text{max}},\text{CNOT}}$.  Once $v_{1}$ has been reached, repeat this process but in the opposite direction, from $i=1$ to $n$, this time commuting backwards instead of forwards.}

\add{The output circuit could then be further optimized using methods that preserve the allowed two-qubit gates, based on templates, for example, or those that dynamically re-allocate physical and logical qubits \cite{li2018tackling, 10.1145/3297858.3304075, finigan2018qubit}.  Of course, our method works best when the segments consisting of CNOT gates are large (many important circuits, such as quantum Fourier transform, fall under this category, increasingly so as the number of qubits increases \cite{nam18}).  The results are shown in Figure 14 compared to IBM's Qiskit transpiler and demonstrate that for input circuits with $\lessapprox 0.05-0.07$ Hadamard to $\text{CNOT}$ ratio, this method consistently outperforms the Qiskit compiler, while it does increasingly worse as the ratio grows larger. The input circuits have 1000 gates and do not take into account connectivity. To generate the initial circuit, each gate is chosen from the set $\{\text{CNOT},S,T,S^{\dag},T^{\dag},H\}$ with respective probabilities $\{p_{\text{CNOT}},p_{S},p_{T},p_{S^{\dag}},p_{T^{\dag}},p_{H}\}$.  For this experiment, $p_{S} = p_{T} = p_{S^{\dag}} = p_{T^{\dag}} = 0.01$, $p_{H} \in [0,0.2]$, and $p_{\text{CNOT}} = 1-\sum_{i \in \{S,T,S^{\dag},T^{\dag},H\}} p_{i}$.}

\begin{figure}[t]
	\includegraphics[width=0.5\textwidth]{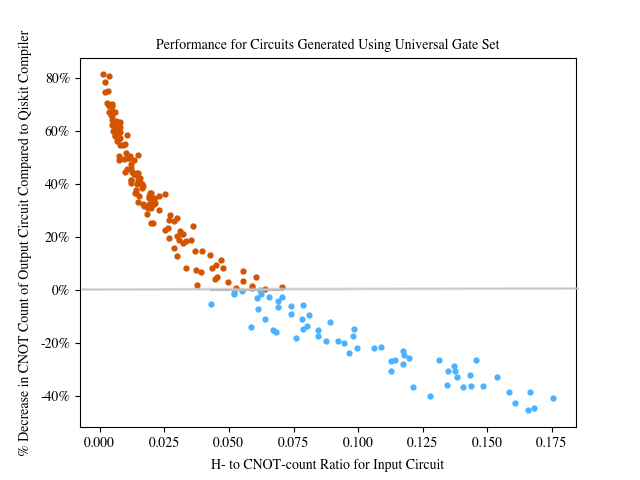}
	\caption{\add{Using the method described in section~\ref{sct:ugs}, then by generating circuits from the set $\mathbb{U}$ with varied Hadamard to CNOT gate ratios, we can clearly see that while the method works well compared to IBM's compiler for circuits that are mainly composed of CNOT gates, its performance decreases as the ratio of Hadamard to CNOT gates increases. Red solid circles represent data points in which our scheme over-performs Qiskit, whereas the blue solid circles represent data points in which our scheme performs worse than Qiskit. The ``sweet-spot" seem to be around 5\%-7\% Hadamard to CNOT ratio.}}
	\label{fgr14}
\end{figure}


\section{Conclusions and open questions\label{sct:conclusions}}
We constructed a circuit synthesis scheme that takes as input the qubit connectivity graph and a quantum circuit over the gate set generated by $\{\text{CNOT},R_{Z}\}$ and outputs a circuit that respects the connectivity of the device. We applied our techniques to Google's Bristlecone 72-qubit quantum chip connectivity, IBM's Tokyo 20-qubit quantum chip connectivity, and Rigetti's Acorn 19-qubit quantum chip connectivity. We also compared the performance of our methods as a function of sparseness of random quantum circuits of 20 qubits.

Since in practice different CNOT gates can be affected by different amount of noise, an idea for further work would be to weight the edges based on the error rate of the CNOT gate between the respective qubits. Techniques adapted from~\cite{6899791} could potentially be integrated into the methods given in this paper to produce circuits with lower error rates. In addition, we plan to improve on efficiency and quality of the results produced by our scheme.

We show how the method can be applied to broader classes of circuits comprising of universal set of gates, however our scheme performs (as expected) less optimally than compilers designed specifically for that task, while over-performing the latter for circuits comprising of $\{\text{CNOT},R_{Z}\}$.

\begin{acknowledgments}
We thank Matt Amy for very useful comments and discussions. We acknowledge support from NSERC and CIFAR. IQC is supported in part by the Government of Canada and the Province of Ontario.

We also want to thank the anonymous referees for the their very helpful suggestions for improving the quality of this manuscript.
\end{acknowledgments}

\bibliographystyle{apsrev4-1}

\begin{thebibliography}{17}%
\makeatletter
\providecommand \@ifxundefined [1]{%
 \@ifx{#1\undefined}
}%
\providecommand \@ifnum [1]{%
 \ifnum #1\expandafter \@firstoftwo
 \else \expandafter \@secondoftwo
 \fi
}%
\providecommand \@ifx [1]{%
 \ifx #1\expandafter \@firstoftwo
 \else \expandafter \@secondoftwo
 \fi
}%
\providecommand \natexlab [1]{#1}%
\providecommand \enquote  [1]{``#1''}%
\providecommand \bibnamefont  [1]{#1}%
\providecommand \bibfnamefont [1]{#1}%
\providecommand \citenamefont [1]{#1}%
\providecommand \href@noop [0]{\@secondoftwo}%
\providecommand \href [0]{\begingroup \@sanitize@url \@href}%
\providecommand \@href[1]{\@@startlink{#1}\@@href}%
\providecommand \@@href[1]{\endgroup#1\@@endlink}%
\providecommand \@sanitize@url [0]{\catcode `\\12\catcode `\$12\catcode
  `\&12\catcode `\#12\catcode `\^12\catcode `\_12\catcode `\%12\relax}%
\providecommand \@@startlink[1]{}%
\providecommand \@@endlink[0]{}%
\providecommand \url  [0]{\begingroup\@sanitize@url \@url }%
\providecommand \@url [1]{\endgroup\@href {#1}{\urlprefix }}%
\providecommand \urlprefix  [0]{URL }%
\providecommand \Eprint [0]{\href }%
\providecommand \doibase [0]{http://dx.doi.org/}%
\providecommand \selectlanguage [0]{\@gobble}%
\providecommand \bibinfo  [0]{\@secondoftwo}%
\providecommand \bibfield  [0]{\@secondoftwo}%
\providecommand \translation [1]{[#1]}%
\providecommand \BibitemOpen [0]{}%
\providecommand \bibitemStop [0]{}%
\providecommand \bibitemNoStop [0]{.\EOS\space}%
\providecommand \EOS [0]{\spacefactor3000\relax}%
\providecommand \BibitemShut  [1]{\csname bibitem#1\endcsname}%
\let\auto@bib@innerbib\@empty
\bibitem [{\citenamefont {Preskill}(2018)}]{Preskill2018quantumcomputingin}%
  \BibitemOpen
  \bibfield  {author} {\bibinfo {author} {\bibfnamefont {J.}~\bibnamefont
  {Preskill}},\ }\href {\doibase 10.22331/q-2018-08-06-79} {\bibfield
  {journal} {\bibinfo  {journal} {{Quantum}}\ }\textbf {\bibinfo {volume}
  {2}},\ \bibinfo {pages} {79} (\bibinfo {year} {2018})}\BibitemShut {NoStop}%
\bibitem [{\citenamefont {Herr}\ \emph {et~al.}(2017)\citenamefont {Herr},
  \citenamefont {Franco},\ and\ \citenamefont {Devitt}}]{herr}%
  \BibitemOpen
  \bibfield  {author} {\bibinfo {author} {\bibfnamefont {D.}~\bibnamefont
  {Herr}}, \bibinfo {author} {\bibfnamefont {N.}~\bibnamefont {Franco}}, \ and\
  \bibinfo {author} {\bibfnamefont {S.}~\bibnamefont {Devitt}},\ }\href@noop {}
  {\bibfield  {journal} {\bibinfo  {journal} {npj Quantum Information}\
  }\textbf {\bibinfo {volume} {3}} (\bibinfo {year} {2017})}\BibitemShut
  {NoStop}%
\bibitem [{\citenamefont {Patel}\ \emph {et~al.}(2008)\citenamefont {Patel},
  \citenamefont {Markov},\ and\ \citenamefont {Hayes}}]{patel}%
  \BibitemOpen
  \bibfield  {author} {\bibinfo {author} {\bibfnamefont {K.}~\bibnamefont
  {Patel}}, \bibinfo {author} {\bibfnamefont {I.}~\bibnamefont {Markov}}, \
  and\ \bibinfo {author} {\bibfnamefont {J.}~\bibnamefont {Hayes}},\
  }\href@noop {} {\bibfield  {journal} {\bibinfo  {journal} {Quantum
  Information and Computation}\ }\textbf {\bibinfo {volume} {8}},\ \bibinfo
  {pages} {282} (\bibinfo {year} {2008})}\BibitemShut {NoStop}%
\bibitem [{\citenamefont {Amy}\ \emph {et~al.}(2014)\citenamefont {Amy},
  \citenamefont {Maslov},\ and\ \citenamefont {Mosca}}]{6899791}%
  \BibitemOpen
  \bibfield  {author} {\bibinfo {author} {\bibfnamefont {M.}~\bibnamefont
  {Amy}}, \bibinfo {author} {\bibfnamefont {D.}~\bibnamefont {Maslov}}, \ and\
  \bibinfo {author} {\bibfnamefont {M.}~\bibnamefont {Mosca}},\ }\href
  {\doibase 10.1109/TCAD.2014.2341953} {\bibfield  {journal} {\bibinfo
  {journal} {Computer-Aided Design of Integrated Circuits and Systems, IEEE
  Transactions on}\ }\textbf {\bibinfo {volume} {33}},\ \bibinfo {pages} {1476}
  (\bibinfo {year} {2014})}\BibitemShut {NoStop}%
\bibitem [{\citenamefont {Abraham}\ \emph {et~al.}(2019)\citenamefont
  {Abraham}, \citenamefont {Akhalwaya}, \citenamefont {Aleksandrowicz},
  \citenamefont {Alexander}, \citenamefont {Alexandrowics}, \citenamefont
  {Arbel}, \citenamefont {Asfaw}, \citenamefont {Azaustre}, \citenamefont
  {Barkoutsos}, \citenamefont {Barron}, \citenamefont {Bello}, \citenamefont
  {Ben-Haim}, \citenamefont {Bevenius}, \citenamefont {Bishop}, \citenamefont
  {Bosch}, \citenamefont {Bucher}, \citenamefont {CZ}, \citenamefont {Cabrera},
  \citenamefont {Calpin}, \citenamefont {Capelluto}, \citenamefont {Carballo},
  \citenamefont {Carrascal}, \citenamefont {Chen}, \citenamefont {Chen},
  \citenamefont {Chen}, \citenamefont {Chow}, \citenamefont {Claus},
  \citenamefont {Clauss}, \citenamefont {Cross}, \citenamefont {Cross},
  \citenamefont {Cruz-Benito}, \citenamefont {Cryoris}, \citenamefont {Culver},
  \citenamefont {C{\'o}rcoles-Gonzales}, \citenamefont {Dague}, \citenamefont
  {Dartiailh}, \citenamefont {Davila}, \citenamefont {Ding}, \citenamefont
  {Dumitrescu}, \citenamefont {Dumon}, \citenamefont {Duran}, \citenamefont
  {Eendebak}, \citenamefont {Egger}, \citenamefont {Everitt}, \citenamefont
  {Fern{\'a}ndez}, \citenamefont {Frisch}, \citenamefont {Fuhrer},
  \citenamefont {GOULD}, \citenamefont {Gacon}, \citenamefont {Gadi},
  \citenamefont {Gago}, \citenamefont {Gambetta}, \citenamefont {Garcia},
  \citenamefont {Garion}, \citenamefont {Gawel-Kus}, \citenamefont
  {Gomez-Mosquera}, \citenamefont {de~la Puente~Gonz{\'a}lez}, \citenamefont
  {Greenberg}, \citenamefont {Gunnels}, \citenamefont {Haide}, \citenamefont
  {Hamamura}, \citenamefont {Havlicek}, \citenamefont {Hellmers}, \citenamefont
  {Herok}, \citenamefont {Horii}, \citenamefont {Howington}, \citenamefont
  {Hu}, \citenamefont {Hu}, \citenamefont {Imai}, \citenamefont {Imamichi},
  \citenamefont {Iten}, \citenamefont {Itoko}, \citenamefont {Javadi-Abhari},
  \citenamefont {Jessica}, \citenamefont {Johns}, \citenamefont {Kanazawa},
  \citenamefont {Karazeev}, \citenamefont {Kassebaum}, \citenamefont
  {Kovyrshin}, \citenamefont {Krishnan}, \citenamefont {Krsulich},
  \citenamefont {Kus}, \citenamefont {LaRose}, \citenamefont {Lambert},
  \citenamefont {Latone}, \citenamefont {Lawrence}, \citenamefont {Liu},
  \citenamefont {Liu}, \citenamefont {Mac}, \citenamefont {Maeng},
  \citenamefont {Malyshev}, \citenamefont {Marecek}, \citenamefont {Marques},
  \citenamefont {Mathews}, \citenamefont {Matsuo}, \citenamefont {McClure},
  \citenamefont {McGarry}, \citenamefont {McKay}, \citenamefont {Meesala},
  \citenamefont {Mezzacapo}, \citenamefont {Midha}, \citenamefont {Minev},
  \citenamefont {Mooring}, \citenamefont {Morales}, \citenamefont {Moran},
  \citenamefont {Murali}, \citenamefont {M{\"u}ggenburg}, \citenamefont
  {Nadlinger}, \citenamefont {Nannicini}, \citenamefont {Nation}, \citenamefont
  {Naveh}, \citenamefont {Nick-Singstock}, \citenamefont {Niroula},
  \citenamefont {Norlen}, \citenamefont {O'Riordan}, \citenamefont
  {Ollitrault}, \citenamefont {Oud}, \citenamefont {Padilha}, \citenamefont
  {Paik}, \citenamefont {Perriello}, \citenamefont {Phan}, \citenamefont
  {Pistoia}, \citenamefont {Pozas-iKerstjens}, \citenamefont {Prutyanov},
  \citenamefont {P{\'e}rez}, \citenamefont {Quintiii}, \citenamefont {Raymond},
  \citenamefont {Redondo}, \citenamefont {Reuter}, \citenamefont
  {Rodr{\'\i}guez}, \citenamefont {Ryu}, \citenamefont {Sandberg},
  \citenamefont {Sathaye}, \citenamefont {Schmitt}, \citenamefont {Schnabel},
  \citenamefont {Scholten}, \citenamefont {Schoute}, \citenamefont {Sertage},
  \citenamefont {Shammah}, \citenamefont {Shi}, \citenamefont {Silva},
  \citenamefont {Siraichi}, \citenamefont {Sivarajah}, \citenamefont {Smolin},
  \citenamefont {Soeken}, \citenamefont {Steenken}, \citenamefont
  {Stypulkoski}, \citenamefont {Takahashi}, \citenamefont {Taylor},
  \citenamefont {Taylour}, \citenamefont {Thomas}, \citenamefont {Tillet},
  \citenamefont {Tod}, \citenamefont {de~la Torre}, \citenamefont {Trabing},
  \citenamefont {Treinish}, \citenamefont {TrishaPe}, \citenamefont {Turner},
  \citenamefont {Vaknin}, \citenamefont {Valcarce}, \citenamefont {Varchon},
  \citenamefont {Vogt-Lee}, \citenamefont {Vuillot}, \citenamefont {Weaver},
  \citenamefont {Wieczorek}, \citenamefont {Wildstrom}, \citenamefont {Wille},
  \citenamefont {Winston}, \citenamefont {Woehr}, \citenamefont {Woerner},
  \citenamefont {Woo}, \citenamefont {Wood}, \citenamefont {Wood},
  \citenamefont {Wood}, \citenamefont {Wootton}, \citenamefont {Yeralin},
  \citenamefont {Yu}, \citenamefont {Zdanski}, \citenamefont {Zoufalc},
  \citenamefont {anedumla}, \citenamefont {azulehner}, \citenamefont
  {bcamorrison}, \citenamefont {brandhsn}, \citenamefont {dennis-liu 1},
  \citenamefont {drholmie}, \citenamefont {elfrocampeador}, \citenamefont
  {fanizzamarco}, \citenamefont {gruu}, \citenamefont {kanejess}, \citenamefont
  {klinvill}, \citenamefont {lerongil}, \citenamefont {ma5x}, \citenamefont
  {merav aharoni}, \citenamefont {mrossinek}, \citenamefont {ordmoj},
  \citenamefont {strickroman}, \citenamefont {tigerjack}, \citenamefont
  {yang.luh},\ and\ \citenamefont {yotamvakninibm}}]{Qiskit}%
  \BibitemOpen
  \bibfield  {author} {\bibinfo {author} {\bibfnamefont {H.}~\bibnamefont
  {Abraham}}, \bibinfo {author} {\bibfnamefont {I.~Y.}\ \bibnamefont
  {Akhalwaya}}, \bibinfo {author} {\bibfnamefont {G.}~\bibnamefont
  {Aleksandrowicz}}, \bibinfo {author} {\bibfnamefont {T.}~\bibnamefont
  {Alexander}}, \bibinfo {author} {\bibfnamefont {G.}~\bibnamefont
  {Alexandrowics}}, \bibinfo {author} {\bibfnamefont {E.}~\bibnamefont
  {Arbel}}, \bibinfo {author} {\bibfnamefont {A.}~\bibnamefont {Asfaw}},
  \bibinfo {author} {\bibfnamefont {C.}~\bibnamefont {Azaustre}}, \bibinfo
  {author} {\bibfnamefont {P.}~\bibnamefont {Barkoutsos}}, \bibinfo {author}
  {\bibfnamefont {G.}~\bibnamefont {Barron}}, \bibinfo {author} {\bibfnamefont
  {L.}~\bibnamefont {Bello}}, \bibinfo {author} {\bibfnamefont
  {Y.}~\bibnamefont {Ben-Haim}}, \bibinfo {author} {\bibfnamefont
  {D.}~\bibnamefont {Bevenius}}, \bibinfo {author} {\bibfnamefont {L.~S.}\
  \bibnamefont {Bishop}}, \bibinfo {author} {\bibfnamefont {S.}~\bibnamefont
  {Bosch}}, \bibinfo {author} {\bibfnamefont {D.}~\bibnamefont {Bucher}},
  \bibinfo {author} {\bibnamefont {CZ}}, \bibinfo {author} {\bibfnamefont
  {F.}~\bibnamefont {Cabrera}}, \bibinfo {author} {\bibfnamefont
  {P.}~\bibnamefont {Calpin}}, \bibinfo {author} {\bibfnamefont
  {L.}~\bibnamefont {Capelluto}}, \bibinfo {author} {\bibfnamefont
  {J.}~\bibnamefont {Carballo}}, \bibinfo {author} {\bibfnamefont
  {G.}~\bibnamefont {Carrascal}}, \bibinfo {author} {\bibfnamefont
  {A.}~\bibnamefont {Chen}}, \bibinfo {author} {\bibfnamefont {C.-F.}\
  \bibnamefont {Chen}}, \bibinfo {author} {\bibfnamefont {R.}~\bibnamefont
  {Chen}}, \bibinfo {author} {\bibfnamefont {J.~M.}\ \bibnamefont {Chow}},
  \bibinfo {author} {\bibfnamefont {C.}~\bibnamefont {Claus}}, \bibinfo
  {author} {\bibfnamefont {C.}~\bibnamefont {Clauss}}, \bibinfo {author}
  {\bibfnamefont {A.~J.}\ \bibnamefont {Cross}}, \bibinfo {author}
  {\bibfnamefont {A.~W.}\ \bibnamefont {Cross}}, \bibinfo {author}
  {\bibfnamefont {J.}~\bibnamefont {Cruz-Benito}}, \bibinfo {author}
  {\bibnamefont {Cryoris}}, \bibinfo {author} {\bibfnamefont {C.}~\bibnamefont
  {Culver}}, \bibinfo {author} {\bibfnamefont {A.~D.}\ \bibnamefont
  {C{\'o}rcoles-Gonzales}}, \bibinfo {author} {\bibfnamefont {S.}~\bibnamefont
  {Dague}}, \bibinfo {author} {\bibfnamefont {M.}~\bibnamefont {Dartiailh}},
  \bibinfo {author} {\bibfnamefont {A.~R.}\ \bibnamefont {Davila}}, \bibinfo
  {author} {\bibfnamefont {D.}~\bibnamefont {Ding}}, \bibinfo {author}
  {\bibfnamefont {E.}~\bibnamefont {Dumitrescu}}, \bibinfo {author}
  {\bibfnamefont {K.}~\bibnamefont {Dumon}}, \bibinfo {author} {\bibfnamefont
  {I.}~\bibnamefont {Duran}}, \bibinfo {author} {\bibfnamefont
  {P.}~\bibnamefont {Eendebak}}, \bibinfo {author} {\bibfnamefont
  {D.}~\bibnamefont {Egger}}, \bibinfo {author} {\bibfnamefont
  {M.}~\bibnamefont {Everitt}}, \bibinfo {author} {\bibfnamefont {P.~M.}\
  \bibnamefont {Fern{\'a}ndez}}, \bibinfo {author} {\bibfnamefont
  {A.}~\bibnamefont {Frisch}}, \bibinfo {author} {\bibfnamefont
  {A.}~\bibnamefont {Fuhrer}}, \bibinfo {author} {\bibfnamefont
  {I.}~\bibnamefont {GOULD}}, \bibinfo {author} {\bibfnamefont
  {J.}~\bibnamefont {Gacon}}, \bibinfo {author} {\bibnamefont {Gadi}}, \bibinfo
  {author} {\bibfnamefont {B.~G.}\ \bibnamefont {Gago}}, \bibinfo {author}
  {\bibfnamefont {J.~M.}\ \bibnamefont {Gambetta}}, \bibinfo {author}
  {\bibfnamefont {L.}~\bibnamefont {Garcia}}, \bibinfo {author} {\bibfnamefont
  {S.}~\bibnamefont {Garion}}, \bibinfo {author} {\bibnamefont {Gawel-Kus}},
  \bibinfo {author} {\bibfnamefont {J.}~\bibnamefont {Gomez-Mosquera}},
  \bibinfo {author} {\bibfnamefont {S.}~\bibnamefont {de~la
  Puente~Gonz{\'a}lez}}, \bibinfo {author} {\bibfnamefont {D.}~\bibnamefont
  {Greenberg}}, \bibinfo {author} {\bibfnamefont {J.~A.}\ \bibnamefont
  {Gunnels}}, \bibinfo {author} {\bibfnamefont {I.}~\bibnamefont {Haide}},
  \bibinfo {author} {\bibfnamefont {I.}~\bibnamefont {Hamamura}}, \bibinfo
  {author} {\bibfnamefont {V.}~\bibnamefont {Havlicek}}, \bibinfo {author}
  {\bibfnamefont {J.}~\bibnamefont {Hellmers}}, \bibinfo {author}
  {\bibfnamefont {{\L}.}~\bibnamefont {Herok}}, \bibinfo {author}
  {\bibfnamefont {H.}~\bibnamefont {Horii}}, \bibinfo {author} {\bibfnamefont
  {C.}~\bibnamefont {Howington}}, \bibinfo {author} {\bibfnamefont
  {S.}~\bibnamefont {Hu}}, \bibinfo {author} {\bibfnamefont {W.}~\bibnamefont
  {Hu}}, \bibinfo {author} {\bibfnamefont {H.}~\bibnamefont {Imai}}, \bibinfo
  {author} {\bibfnamefont {T.}~\bibnamefont {Imamichi}}, \bibinfo {author}
  {\bibfnamefont {R.}~\bibnamefont {Iten}}, \bibinfo {author} {\bibfnamefont
  {T.}~\bibnamefont {Itoko}}, \bibinfo {author} {\bibfnamefont
  {A.}~\bibnamefont {Javadi-Abhari}}, \bibinfo {author} {\bibnamefont
  {Jessica}}, \bibinfo {author} {\bibfnamefont {K.}~\bibnamefont {Johns}},
  \bibinfo {author} {\bibfnamefont {N.}~\bibnamefont {Kanazawa}}, \bibinfo
  {author} {\bibfnamefont {A.}~\bibnamefont {Karazeev}}, \bibinfo {author}
  {\bibfnamefont {P.}~\bibnamefont {Kassebaum}}, \bibinfo {author}
  {\bibfnamefont {A.}~\bibnamefont {Kovyrshin}}, \bibinfo {author}
  {\bibfnamefont {V.}~\bibnamefont {Krishnan}}, \bibinfo {author}
  {\bibfnamefont {K.}~\bibnamefont {Krsulich}}, \bibinfo {author}
  {\bibfnamefont {G.}~\bibnamefont {Kus}}, \bibinfo {author} {\bibfnamefont
  {R.}~\bibnamefont {LaRose}}, \bibinfo {author} {\bibfnamefont
  {R.}~\bibnamefont {Lambert}}, \bibinfo {author} {\bibfnamefont
  {J.}~\bibnamefont {Latone}}, \bibinfo {author} {\bibfnamefont
  {S.}~\bibnamefont {Lawrence}}, \bibinfo {author} {\bibfnamefont
  {D.}~\bibnamefont {Liu}}, \bibinfo {author} {\bibfnamefont {P.}~\bibnamefont
  {Liu}}, \bibinfo {author} {\bibfnamefont {P.~B.~Z.}\ \bibnamefont {Mac}},
  \bibinfo {author} {\bibfnamefont {Y.}~\bibnamefont {Maeng}}, \bibinfo
  {author} {\bibfnamefont {A.}~\bibnamefont {Malyshev}}, \bibinfo {author}
  {\bibfnamefont {J.}~\bibnamefont {Marecek}}, \bibinfo {author} {\bibfnamefont
  {M.}~\bibnamefont {Marques}}, \bibinfo {author} {\bibfnamefont
  {D.}~\bibnamefont {Mathews}}, \bibinfo {author} {\bibfnamefont
  {A.}~\bibnamefont {Matsuo}}, \bibinfo {author} {\bibfnamefont {D.~T.}\
  \bibnamefont {McClure}}, \bibinfo {author} {\bibfnamefont {C.}~\bibnamefont
  {McGarry}}, \bibinfo {author} {\bibfnamefont {D.}~\bibnamefont {McKay}},
  \bibinfo {author} {\bibfnamefont {S.}~\bibnamefont {Meesala}}, \bibinfo
  {author} {\bibfnamefont {A.}~\bibnamefont {Mezzacapo}}, \bibinfo {author}
  {\bibfnamefont {R.}~\bibnamefont {Midha}}, \bibinfo {author} {\bibfnamefont
  {Z.}~\bibnamefont {Minev}}, \bibinfo {author} {\bibfnamefont {M.~D.}\
  \bibnamefont {Mooring}}, \bibinfo {author} {\bibfnamefont {R.}~\bibnamefont
  {Morales}}, \bibinfo {author} {\bibfnamefont {N.}~\bibnamefont {Moran}},
  \bibinfo {author} {\bibfnamefont {P.}~\bibnamefont {Murali}}, \bibinfo
  {author} {\bibfnamefont {J.}~\bibnamefont {M{\"u}ggenburg}}, \bibinfo
  {author} {\bibfnamefont {D.}~\bibnamefont {Nadlinger}}, \bibinfo {author}
  {\bibfnamefont {G.}~\bibnamefont {Nannicini}}, \bibinfo {author}
  {\bibfnamefont {P.}~\bibnamefont {Nation}}, \bibinfo {author} {\bibfnamefont
  {Y.}~\bibnamefont {Naveh}}, \bibinfo {author} {\bibnamefont
  {Nick-Singstock}}, \bibinfo {author} {\bibfnamefont {P.}~\bibnamefont
  {Niroula}}, \bibinfo {author} {\bibfnamefont {H.}~\bibnamefont {Norlen}},
  \bibinfo {author} {\bibfnamefont {L.~J.}\ \bibnamefont {O'Riordan}}, \bibinfo
  {author} {\bibfnamefont {P.}~\bibnamefont {Ollitrault}}, \bibinfo {author}
  {\bibfnamefont {S.}~\bibnamefont {Oud}}, \bibinfo {author} {\bibfnamefont
  {D.}~\bibnamefont {Padilha}}, \bibinfo {author} {\bibfnamefont
  {H.}~\bibnamefont {Paik}}, \bibinfo {author} {\bibfnamefont {S.}~\bibnamefont
  {Perriello}}, \bibinfo {author} {\bibfnamefont {A.}~\bibnamefont {Phan}},
  \bibinfo {author} {\bibfnamefont {M.}~\bibnamefont {Pistoia}}, \bibinfo
  {author} {\bibfnamefont {A.}~\bibnamefont {Pozas-iKerstjens}}, \bibinfo
  {author} {\bibfnamefont {V.}~\bibnamefont {Prutyanov}}, \bibinfo {author}
  {\bibfnamefont {J.}~\bibnamefont {P{\'e}rez}}, \bibinfo {author}
  {\bibnamefont {Quintiii}}, \bibinfo {author} {\bibfnamefont {R.}~\bibnamefont
  {Raymond}}, \bibinfo {author} {\bibfnamefont {R.~M.-C.}\ \bibnamefont
  {Redondo}}, \bibinfo {author} {\bibfnamefont {M.}~\bibnamefont {Reuter}},
  \bibinfo {author} {\bibfnamefont {D.~M.}\ \bibnamefont {Rodr{\'\i}guez}},
  \bibinfo {author} {\bibfnamefont {M.}~\bibnamefont {Ryu}}, \bibinfo {author}
  {\bibfnamefont {M.}~\bibnamefont {Sandberg}}, \bibinfo {author}
  {\bibfnamefont {N.}~\bibnamefont {Sathaye}}, \bibinfo {author} {\bibfnamefont
  {B.}~\bibnamefont {Schmitt}}, \bibinfo {author} {\bibfnamefont
  {C.}~\bibnamefont {Schnabel}}, \bibinfo {author} {\bibfnamefont {T.~L.}\
  \bibnamefont {Scholten}}, \bibinfo {author} {\bibfnamefont {E.}~\bibnamefont
  {Schoute}}, \bibinfo {author} {\bibfnamefont {I.~F.}\ \bibnamefont
  {Sertage}}, \bibinfo {author} {\bibfnamefont {N.}~\bibnamefont {Shammah}},
  \bibinfo {author} {\bibfnamefont {Y.}~\bibnamefont {Shi}}, \bibinfo {author}
  {\bibfnamefont {A.}~\bibnamefont {Silva}}, \bibinfo {author} {\bibfnamefont
  {Y.}~\bibnamefont {Siraichi}}, \bibinfo {author} {\bibfnamefont
  {S.}~\bibnamefont {Sivarajah}}, \bibinfo {author} {\bibfnamefont {J.~A.}\
  \bibnamefont {Smolin}}, \bibinfo {author} {\bibfnamefont {M.}~\bibnamefont
  {Soeken}}, \bibinfo {author} {\bibfnamefont {D.}~\bibnamefont {Steenken}},
  \bibinfo {author} {\bibfnamefont {M.}~\bibnamefont {Stypulkoski}}, \bibinfo
  {author} {\bibfnamefont {H.}~\bibnamefont {Takahashi}}, \bibinfo {author}
  {\bibfnamefont {C.}~\bibnamefont {Taylor}}, \bibinfo {author} {\bibfnamefont
  {P.}~\bibnamefont {Taylour}}, \bibinfo {author} {\bibfnamefont
  {S.}~\bibnamefont {Thomas}}, \bibinfo {author} {\bibfnamefont
  {M.}~\bibnamefont {Tillet}}, \bibinfo {author} {\bibfnamefont
  {M.}~\bibnamefont {Tod}}, \bibinfo {author} {\bibfnamefont {E.}~\bibnamefont
  {de~la Torre}}, \bibinfo {author} {\bibfnamefont {K.}~\bibnamefont
  {Trabing}}, \bibinfo {author} {\bibfnamefont {M.}~\bibnamefont {Treinish}},
  \bibinfo {author} {\bibnamefont {TrishaPe}}, \bibinfo {author} {\bibfnamefont
  {W.}~\bibnamefont {Turner}}, \bibinfo {author} {\bibfnamefont
  {Y.}~\bibnamefont {Vaknin}}, \bibinfo {author} {\bibfnamefont {C.~R.}\
  \bibnamefont {Valcarce}}, \bibinfo {author} {\bibfnamefont {F.}~\bibnamefont
  {Varchon}}, \bibinfo {author} {\bibfnamefont {D.}~\bibnamefont {Vogt-Lee}},
  \bibinfo {author} {\bibfnamefont {C.}~\bibnamefont {Vuillot}}, \bibinfo
  {author} {\bibfnamefont {J.}~\bibnamefont {Weaver}}, \bibinfo {author}
  {\bibfnamefont {R.}~\bibnamefont {Wieczorek}}, \bibinfo {author}
  {\bibfnamefont {J.~A.}\ \bibnamefont {Wildstrom}}, \bibinfo {author}
  {\bibfnamefont {R.}~\bibnamefont {Wille}}, \bibinfo {author} {\bibfnamefont
  {E.}~\bibnamefont {Winston}}, \bibinfo {author} {\bibfnamefont {J.~J.}\
  \bibnamefont {Woehr}}, \bibinfo {author} {\bibfnamefont {S.}~\bibnamefont
  {Woerner}}, \bibinfo {author} {\bibfnamefont {R.}~\bibnamefont {Woo}},
  \bibinfo {author} {\bibfnamefont {C.~J.}\ \bibnamefont {Wood}}, \bibinfo
  {author} {\bibfnamefont {R.}~\bibnamefont {Wood}}, \bibinfo {author}
  {\bibfnamefont {S.}~\bibnamefont {Wood}}, \bibinfo {author} {\bibfnamefont
  {J.}~\bibnamefont {Wootton}}, \bibinfo {author} {\bibfnamefont
  {D.}~\bibnamefont {Yeralin}}, \bibinfo {author} {\bibfnamefont
  {J.}~\bibnamefont {Yu}}, \bibinfo {author} {\bibfnamefont {L.}~\bibnamefont
  {Zdanski}}, \bibinfo {author} {\bibnamefont {Zoufalc}}, \bibinfo {author}
  {\bibnamefont {anedumla}}, \bibinfo {author} {\bibnamefont {azulehner}},
  \bibinfo {author} {\bibnamefont {bcamorrison}}, \bibinfo {author}
  {\bibnamefont {brandhsn}}, \bibinfo {author} {\bibnamefont {dennis-liu 1}},
  \bibinfo {author} {\bibnamefont {drholmie}}, \bibinfo {author} {\bibnamefont
  {elfrocampeador}}, \bibinfo {author} {\bibnamefont {fanizzamarco}}, \bibinfo
  {author} {\bibnamefont {gruu}}, \bibinfo {author} {\bibnamefont {kanejess}},
  \bibinfo {author} {\bibnamefont {klinvill}}, \bibinfo {author} {\bibnamefont
  {lerongil}}, \bibinfo {author} {\bibnamefont {ma5x}}, \bibinfo {author}
  {\bibnamefont {merav aharoni}}, \bibinfo {author} {\bibnamefont {mrossinek}},
  \bibinfo {author} {\bibnamefont {ordmoj}}, \bibinfo {author} {\bibnamefont
  {strickroman}}, \bibinfo {author} {\bibnamefont {tigerjack}}, \bibinfo
  {author} {\bibnamefont {yang.luh}}, \ and\ \bibinfo {author} {\bibnamefont
  {yotamvakninibm}},\ }\href {\doibase 10.5281/zenodo.2562110} {\enquote
  {\bibinfo {title} {Qiskit: An open-source framework for quantum computing},}\
  } (\bibinfo {year} {2019})\BibitemShut {NoStop}%
\bibitem [{\citenamefont {Kissinger}\ and\ \citenamefont {van~de
  Griend}()}]{1904.00633}%
  \BibitemOpen
  \bibfield  {author} {\bibinfo {author} {\bibfnamefont {A.}~\bibnamefont
  {Kissinger}}\ and\ \bibinfo {author} {\bibfnamefont {A.~M.}\ \bibnamefont
  {van~de Griend}},\ }\href@noop {} {\enquote {\bibinfo {title} {Cnot circuit
  extraction for topologically-constrained quantum memories},}\ }\bibinfo
  {note} {ArXiv:1904.00633}\BibitemShut {NoStop}%
\bibitem [{\citenamefont {Mosca}\ \emph {et~al.}(2019)\citenamefont {Mosca},
  \citenamefont {Roetteler},\ and\ \citenamefont
  {Selinger}}]{mosca_et_al:DR:2019:10329}%
  \BibitemOpen
  \bibfield  {author} {\bibinfo {author} {\bibfnamefont {M.}~\bibnamefont
  {Mosca}}, \bibinfo {author} {\bibfnamefont {M.}~\bibnamefont {Roetteler}}, \
  and\ \bibinfo {author} {\bibfnamefont {P.}~\bibnamefont {Selinger}},\ }\href
  {\doibase 10.4230/DagRep.8.9.112} {\bibfield  {journal} {\bibinfo  {journal}
  {Dagstuhl Reports}\ }\textbf {\bibinfo {volume} {8}},\ \bibinfo {pages} {112}
  (\bibinfo {year} {2019})}\BibitemShut {NoStop}%
\bibitem [{\citenamefont {Shende}\ \emph {et~al.}(2003)\citenamefont {Shende},
  \citenamefont {Prasad}, \citenamefont {Markov},\ and\ \citenamefont
  {Hayes}}]{shende}%
  \BibitemOpen
  \bibfield  {author} {\bibinfo {author} {\bibfnamefont {V.}~\bibnamefont
  {Shende}}, \bibinfo {author} {\bibfnamefont {A.}~\bibnamefont {Prasad}},
  \bibinfo {author} {\bibfnamefont {I.}~\bibnamefont {Markov}}, \ and\ \bibinfo
  {author} {\bibfnamefont {J.}~\bibnamefont {Hayes}},\ }\href@noop {}
  {\bibfield  {journal} {\bibinfo  {journal} {IEEE Trans. on CAD}\ ,\ \bibinfo
  {pages} {710}} (\bibinfo {year} {2003})}\BibitemShut {NoStop}%
\bibitem [{\citenamefont {Byrka}\ \emph {et~al.}(2010)\citenamefont {Byrka},
  \citenamefont {Grandoni}, \citenamefont {Rothvo{\ss}},\ and\ \citenamefont
  {Sanita}}]{byrka}%
  \BibitemOpen
  \bibfield  {author} {\bibinfo {author} {\bibfnamefont {J.}~\bibnamefont
  {Byrka}}, \bibinfo {author} {\bibfnamefont {F.}~\bibnamefont {Grandoni}},
  \bibinfo {author} {\bibfnamefont {T.}~\bibnamefont {Rothvo{\ss}}}, \ and\
  \bibinfo {author} {\bibfnamefont {L.}~\bibnamefont {Sanita}},\ }\href@noop {}
  {\bibfield  {journal} {\bibinfo  {journal} {Proc. of STOC}\ ,\ \bibinfo
  {pages} {583}} (\bibinfo {year} {2010})}\BibitemShut {NoStop}%
\bibitem [{\citenamefont {Robins}\ and\ \citenamefont
  {Zelikovsky}(2000)}]{robins}%
  \BibitemOpen
  \bibfield  {author} {\bibinfo {author} {\bibfnamefont {G.}~\bibnamefont
  {Robins}}\ and\ \bibinfo {author} {\bibfnamefont {A.}~\bibnamefont
  {Zelikovsky}},\ }\href@noop {} {\bibfield  {journal} {\bibinfo  {journal}
  {7th Annual ACM-SIAM Symposium on Discrete Algorithms}\ } (\bibinfo {year}
  {2000})}\BibitemShut {NoStop}%
\bibitem [{\citenamefont {Wang}(1985)}]{wang}%
  \BibitemOpen
  \bibfield  {author} {\bibinfo {author} {\bibfnamefont {S.}~\bibnamefont
  {Wang}},\ }\href@noop {} {\bibfield  {journal} {\bibinfo  {journal} {Proc.
  International Workshop on Graphtheoretic Concepts in Computer Science}\ ,\
  \bibinfo {pages} {387}} (\bibinfo {year} {1985})}\BibitemShut {NoStop}%
\bibitem [{\citenamefont {Hwang}\ and\ \citenamefont {Richards}(1992)}]{hwang}%
  \BibitemOpen
  \bibfield  {author} {\bibinfo {author} {\bibfnamefont {F.}~\bibnamefont
  {Hwang}}\ and\ \bibinfo {author} {\bibfnamefont {K.}~\bibnamefont
  {Richards}},\ }\href@noop {} {\bibfield  {journal} {\bibinfo  {journal}
  {Networks}\ }\textbf {\bibinfo {volume} {22}},\ \bibinfo {pages} {55}
  (\bibinfo {year} {1992})}\BibitemShut {NoStop}%
\bibitem [{abd(2014)}]{abdessaied14}%
  \BibitemOpen
  \href@noop {} {\bibfield  {journal} {\bibinfo  {journal} {International
  Conference on Reversible Computation, Lecture Notes in Computer Science}\
  }\textbf {\bibinfo {volume} {8507}},\ \bibinfo {pages} {149} (\bibinfo {year}
  {2014})}\BibitemShut {NoStop}%
\bibitem [{\citenamefont {Li}\ \emph {et~al.}(2018)\citenamefont {Li},
  \citenamefont {Ding},\ and\ \citenamefont {Xie}}]{li2018tackling}%
  \BibitemOpen
  \bibfield  {author} {\bibinfo {author} {\bibfnamefont {G.}~\bibnamefont
  {Li}}, \bibinfo {author} {\bibfnamefont {Y.}~\bibnamefont {Ding}}, \ and\
  \bibinfo {author} {\bibfnamefont {Y.}~\bibnamefont {Xie}},\ }\href@noop {}
  {\enquote {\bibinfo {title} {Tackling the qubit mapping problem for nisq-era
  quantum devices},}\ } (\bibinfo {year} {2018}),\ \Eprint
  {http://arxiv.org/abs/1809.02573} {arXiv:1809.02573 [cs.ET]} \BibitemShut
  {NoStop}%
\bibitem [{\citenamefont {Murali}\ \emph {et~al.}(2019)\citenamefont {Murali},
  \citenamefont {Baker}, \citenamefont {Javadi-Abhari}, \citenamefont {Chong},\
  and\ \citenamefont {Martonosi}}]{10.1145/3297858.3304075}%
  \BibitemOpen
  \bibfield  {author} {\bibinfo {author} {\bibfnamefont {P.}~\bibnamefont
  {Murali}}, \bibinfo {author} {\bibfnamefont {J.~M.}\ \bibnamefont {Baker}},
  \bibinfo {author} {\bibfnamefont {A.}~\bibnamefont {Javadi-Abhari}}, \bibinfo
  {author} {\bibfnamefont {F.~T.}\ \bibnamefont {Chong}}, \ and\ \bibinfo
  {author} {\bibfnamefont {M.}~\bibnamefont {Martonosi}},\ }in\ \href {\doibase
  10.1145/3297858.3304075} {\emph {\bibinfo {booktitle} {Proceedings of the
  Twenty-Fourth International Conference on Architectural Support for
  Programming Languages and Operating Systems}}},\ \bibinfo {series and number}
  {ASPLOS ’19}\ (\bibinfo  {publisher} {Association for Computing
  Machinery},\ \bibinfo {address} {New York, NY, USA},\ \bibinfo {year}
  {2019})\ p.\ \bibinfo {pages} {1015–1029}\BibitemShut {NoStop}%
\bibitem [{\citenamefont {Finigan}\ \emph {et~al.}(2018)\citenamefont
  {Finigan}, \citenamefont {Cubeddu}, \citenamefont {Lively}, \citenamefont
  {Flick},\ and\ \citenamefont {Narang}}]{finigan2018qubit}%
  \BibitemOpen
  \bibfield  {author} {\bibinfo {author} {\bibfnamefont {W.}~\bibnamefont
  {Finigan}}, \bibinfo {author} {\bibfnamefont {M.}~\bibnamefont {Cubeddu}},
  \bibinfo {author} {\bibfnamefont {T.}~\bibnamefont {Lively}}, \bibinfo
  {author} {\bibfnamefont {J.}~\bibnamefont {Flick}}, \ and\ \bibinfo {author}
  {\bibfnamefont {P.}~\bibnamefont {Narang}},\ }\href@noop {} {\enquote
  {\bibinfo {title} {Qubit allocation for noisy intermediate-scale quantum
  computers},}\ } (\bibinfo {year} {2018}),\ \Eprint
  {http://arxiv.org/abs/1810.08291} {arXiv:1810.08291 [quant-ph]} \BibitemShut
  {NoStop}%
\bibitem [{\citenamefont {Nam}\ \emph {et~al.}(2018)\citenamefont {Nam},
  \citenamefont {Ross}, \citenamefont {Su}, \citenamefont {Childs},\ and\
  \citenamefont {Maslov}}]{nam18}%
  \BibitemOpen
  \bibfield  {author} {\bibinfo {author} {\bibfnamefont {Y.}~\bibnamefont
  {Nam}}, \bibinfo {author} {\bibfnamefont {N.~J.}\ \bibnamefont {Ross}},
  \bibinfo {author} {\bibfnamefont {Y.}~\bibnamefont {Su}}, \bibinfo {author}
  {\bibfnamefont {A.~M.}\ \bibnamefont {Childs}}, \ and\ \bibinfo {author}
  {\bibfnamefont {D.}~\bibnamefont {Maslov}},\ }\href {\doibase
  10.1038/s41534-018-0072-4} {\bibfield  {journal} {\bibinfo  {journal} {npj
  Quantum Information}\ }\textbf {\bibinfo {volume} {4}},\ \bibinfo {pages}
  {23} (\bibinfo {year} {2018})}\BibitemShut {NoStop}%
\end{thebibliography}

\onecolumngrid
\pagebreak
\appendix
\begin{figure*}[h]
\section{Example\label{apdx:example}}
\begin{tabularx}{\textwidth}{|XX|XX|}
\hline 
\small\textbf{1)} &\hfill &\small\textbf{2)}  & \small\hfill \\
\small$\left( \begin{array} {>{\columncolor{gray!40}}cccccc}
1 & 1 & 0 & 1 & 1 & 0 \\
0 & 0 & 1 & 1 & 0 & 1 \\
1 & 0 & 1 & 0 & 1 & 0 \\
1 & 1 & 0 & 1 & 0 & 0 \\
1 & 1 & 1 & 1 & 0 & 0 \\
0 & 1 & 0 & 1 & 0 & 1 
\end{array} \right)$
&\hfill
\raisebox{-.5\height}{\includegraphics[scale=0.65]{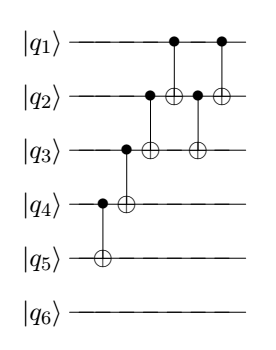}}&
\small$\left( \begin{array} {c>{\columncolor{gray!40}}ccccc}
1 & 1 & 0 & 1 & 1 & 0 \\
0 & 0 & 1 & 1 & 0 & 1 \\
0 & 1 & 1 & 1 & 0 & 0 \\
0 & 1 & 1 & 1 & 1 & 0 \\
0 & 0 & 1 & 0 & 0 & 0 \\
0 & 1 & 0 & 1 & 0 & 1 
\end{array} \right)$
&\hfill\raisebox{-.5\height}{\includegraphics[scale=0.65]{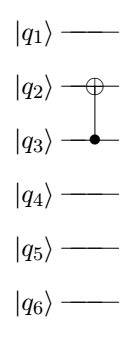}} \\
\hline 
\small\textbf{3)} &\hfill &\small\textbf{4)}  & \small\hfill \\
\small $\left( \begin{array} {c>{\columncolor{gray!40}}ccccc}
1 & 1 & 0 & 1 & 1 & 0 \\
0 & 1 & 0 & 0 & 0 & 1 \\
0 & 1 & 1 & 1 & 0 & 0 \\
0 & 1 & 1 & 1 & 1 & 0 \\
0 & 0 & 1 & 0 & 0 & 0 \\
0 & 1 & 0 & 1 & 0 & 1 
\end{array} \right)$
&\hfill
\raisebox{-.5\height}{\includegraphics[scale=0.65]{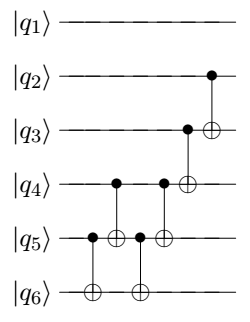}}&
\small $\left( \begin{array} {cc>{\columncolor{gray!40}}cccc}
1 & 1 & 0 & 1 & 1 & 0 \\
0 & 1 & 0 & 0 & 0 & 1 \\
0 & 0 & 1 & 1 & 0 & 1 \\
0 & 0 & 0 & 0 & 1 & 0 \\
0 & 0 & 1 & 0 & 0 & 0 \\
0 & 0 & 1 & 0 & 1 & 1 
\end{array} \right)$
&\hfill
\raisebox{-.5\height}{\includegraphics[scale=0.65]{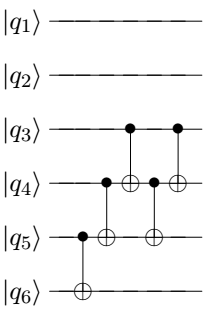}} \\
\hline 
\small\textbf{5)} &\hfill &\small\textbf{6)}  & \small\hfill \\
\small $\left( \begin{array} {ccc>{\columncolor{gray!40}}ccc}
1 & 1 & 0 & 1 & 1 & 0 \\
0 & 1 & 0 & 0 & 0 & 1 \\
0 & 0 & 1 & 1 & 0 & 1 \\
0 & 0 & 0 & 0 & 1 & 0 \\
0 & 0 & 0 & 1 & 0 & 1 \\
0 & 0 & 0 & 0 & 1 & 1 
\end{array} \right)$
&\hfill \raisebox{-.5\height}{\includegraphics[scale=0.65]{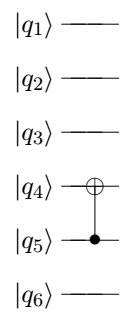}}
&\small$\left( \begin{array} {ccc>{\columncolor{gray!40}}ccc}
1 & 1 & 0 & 1 & 1 & 0 \\
0 & 1 & 0 & 0 & 0 & 1 \\
0 & 0 & 1 & 1 & 0 & 1 \\
0 & 0 & 0 & 1 & 1 & 1 \\
0 & 0 & 0 & 1 & 0 & 1 \\
0 & 0 & 0 & 0 & 1 & 1 
\end{array} \right)$
&\hfill \raisebox{-.5\height}{\includegraphics[scale=0.65]{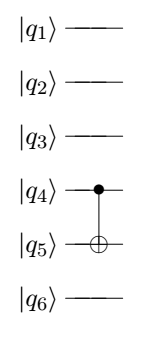}} \\
\hline 
\small\textbf{7)} &\hfill &\small\textbf{8)}  & \small\hfill \\
 $\left( \begin{array} {cccc>{\columncolor{gray!40}}cc}
1 & 1 & 0 & 1 & 1 & 0 \\
0 & 1 & 0 & 0 & 0 & 1 \\
0 & 0 & 1 & 1 & 0 & 1 \\
0 & 0 & 0 & 1 & 1 & 1 \\
0 & 0 & 0 & 0 & 1 & 0 \\
0 & 0 & 0 & 0 & 1 & 1 
\end{array} \right)$
&\hfill \raisebox{-.5\height}{\includegraphics[scale=0.65]{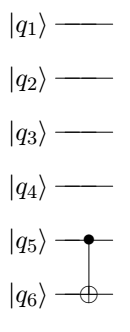}}
&\small $\left( \begin{array} {>{\columncolor{gray!40}}cccccc}
1 & 0 & 0 & 0 & 0 & 0 \\
1 & 1 & 0 & 0 & 0 & 0 \\
0 & 0 & 1 & 0 & 0 & 0 \\
1 & 0 & 1 & 1 & 0 & 0 \\
1 & 0 & 0 & 1 & 1 & 0 \\
0 & 1 & 1 & 1 & 0 & 1 
\end{array} \right)$
&\hfill \raisebox{-.5\height}{\includegraphics[scale=0.65]{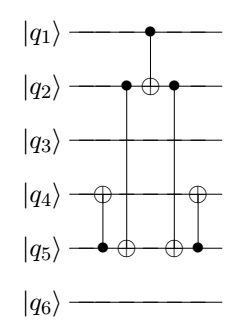}}  \\
\hline 
\small\textbf{9)} &\hfill &\small\textbf{10)}  & \small\hfill \\
\small $\left( \begin{array} {c>{\columncolor{gray!40}}ccccc}
1 & 0 & 0 & 0 & 0 & 0 \\
0 & 1 & 0 & 0 & 0 & 0 \\
0 & 0 & 1 & 0 & 0 & 0 \\
0 & 0 & 1 & 1 & 0 & 0 \\
0 & 0 & 0 & 1 & 1 & 0 \\
0 & 1 & 1 & 1 & 0 & 1 
\end{array} \right)$
&\hfill \raisebox{-.5\height}{\includegraphics[scale=0.65]{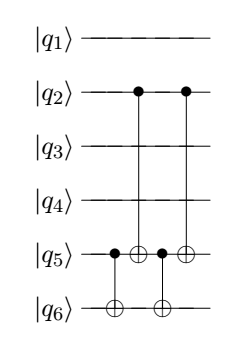}}
&\small $\left( \begin{array} {cc>{\columncolor{gray!40}}cccc}
1 & 0 & 0 & 0 & 0 & 0 \\
0 & 1 & 0 & 0 & 0 & 0 \\
0 & 0 & 1 & 0 & 0 & 0 \\
0 & 0 & 1 & 1 & 0 & 0 \\
0 & 0 & 0 & 1 & 1 & 0 \\
0 & 0 & 1 & 1 & 0 & 1 
\end{array} \right)$
&\hfill \raisebox{-.5\height}{\includegraphics[scale=0.65]{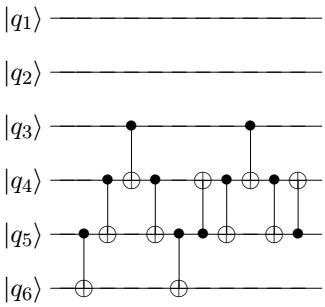}}\\
\hline
\end{tabularx}
\end{figure*}

\begin{figure*}
\begin{tabularx}{\textwidth}{|XX|XX|}
\hline 
\small\textbf{11)} &\hfill &\small\textbf{12)}  & \small\hfill \\
\small $\left( \begin{array} {ccc>{\columncolor{gray!40}}ccc}
1 & 0 & 0 & 0 & 0 & 0 \\
0 & 1 & 0 & 0 & 0 & 0 \\
0 & 0 & 1 & 0 & 0 & 0 \\
0 & 0 & 0 & 1 & 0 & 0 \\
0 & 0 & 0 & 1 & 1 & 0 \\
0 & 0 & 0 & 1 & 0 & 1 
\end{array} \right)$
&\hfill \raisebox{-.5\height}{\includegraphics[scale=0.65]{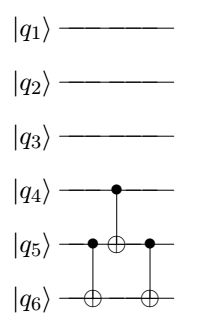}}
&\small $\left( \begin{array} {cccc>{\columncolor{gray!40}}cc}
1 & 0 & 0 & 0 & 0 & 0 \\
0 & 1 & 0 & 0 & 0 & 0 \\
0 & 0 & 1 & 0 & 0 & 0 \\
0 & 0 & 0 & 1 & 0 & 0 \\
0 & 0 & 0 & 0 & 1 & 0 \\
0 & 0 & 0 & 0 & 0 & 1 
\end{array} \right)$
&\hfill \\
\hline
\end{tabularx}
\caption{Worked through example of circuit synthesis producing the linear transformation given Fig.~\ref{fgr3} (a) on connectivity given in Fig.~\ref{fgr3} (b). Fully synthesized circuit can be produced by putting together the gates from steps 8-11 and switching the controls and targets, then putting together the gates from steps 1-7 and reversing them, leaving the controls and targets as shown.}
\label{fgr15}
\end{figure*}

\end{document}